\documentclass[pre,english,10pt]{revtex4}
\usepackage{times}
\usepackage[T1]{fontenc}
\usepackage[latin1]{inputenc}
\usepackage{geometry}
\usepackage{rotating}
\usepackage{color}
\usepackage{graphicx}
\usepackage{setspace}

\usepackage{amsmath}    
\usepackage{amsfonts}
\usepackage{amssymb}	
\usepackage{natbib} 

\newcommand{\dx}{\mathrm{d}x}

\newcommand{\argmax}{\operatorname*{argmax}}

\def\citen#1{\citeauthor{#1} \citeyear{#1}}

\newcommand{\entropy}{\Omega}

\begin{document}

\title{\Large Multiple-line inference of selection on quantitative traits}
\author{N. Riedel$^1$, B. S. Khatri$^2$, M. L\"assig$^1$, and J. Berg$^1$}
\affiliation{$^1$Institut f\"ur Theoretische Physik, University of Cologne - Z\"ulpicher Stra\ss e 77, 50937 K\"oln, Germany
\\ $^2$The Francis Crick Institute, Mill Hill Laboratory, The Ridgeway, London, U.K.}
\maketitle

\section*{\large Abstract}

Trait differences between species may be attributable to natural selection. However, quantifying the strength of evidence for selection acting on a particular trait is a difficult task. Here we develop a population-genetic test for selection acting on a quantitative trait which is based on multiple-line crosses. We show that using multiple lines increases both the power and the scope of selection inference. First, a test based on three or more lines detects selection with strongly increased statistical significance, and we show explicitly how the sensitivity of the test depends on the number of lines. Second, a multiple-line test allows to distinguish different lineage-specific selection scenarios. Our analytical results are complemented by extensive numerical simulations. We then apply the multiple-line test to {QTL} data on floral character traits in plant species of the \textit{Mimulus} genus and on photoperiodic traits in different maize strains, where we find a signatures of lineage-specific selection not seen in a two-line test.

{\centering
\section{\large Introduction}
}

Extensive experimental work has helped reveal the genetic architecture of quantitative traits (\citen{Dilda2002};\,\citen{Mezey2005};\,\citen{Nuzhdin2005};\,\citen{MackayLyman2005};\,\citen{Brem2005};\,\citen{Flint2009}), allowing one to study the basis of trait variation within and across species. A long-term goal of {QTL} research is to understand the mapping from genotype to phenotype underlying a particular quantitative trait. Crosses between individuals from different lines are used to identify loci whose states are statistically correlated with a particular trait.  However, the ability of {QTL} studies to identify the molecular basis of quantitative traits is still limited; it is especially difficult to pinpoint genetic loci influencing a trait (\citen{Mackay2009}). Targeted efforts have been made to resolve loci at the level of single genes or even nucleotides (\citen{Pasyukova2000};\,\citen{Fanara2002};\,\citen{DeLuca2003};\,\citen{Moehring2004};\,\citen{Harbison2004};\,\citen{Jordan2006}), but these cases are still the exception.

In recent years, {QTL} experiments have also been extended to crosses between multiple lines. Harnessing information from several lines drastically increases the power and accuracy of {QTL} identification (\citen{Rebai1993};\,\citen{Steinhoff2011}), allowing one to test for epistatic interactions (\citen{Blanc2006};\,\citen{Jannink2001}) and increasing the genetic variability that can be accessed (\citen{Blanc2006}). For instance, all loci that have the same allele in two lines also have the same allele in all crosses of these lines. In the absence of genetic variance, the effect of such a locus on a trait cannot be determined. Analysing more than two lines increases the number of loci that differ by state in at least one line, allowing one to identify more loci affecting a quantitative trait. Multiple-line pairwise crosses are most common in animal and plant breeding (\citen{Rueckert2010};\,\citen{Blanc2006}), where often many different lines are available for crossing. However, the extension to multiple line crosses also brings new challenges. For instance, choosing the right mating design for the {QTL} experiments is important for multiple-line crosses (\citen{Crepieux2004};\,\citen{Verhoeven2006}). Since most statistical methods for {QTL} identification developed for two-line crosses cannot easily be extended to the multiple-line case, new and more sophisticated methods have been developed (\citen{Xie1998}).  These methods are based on least-squares regression (\citen{Rebai2000}), maximum likelihood (\citen{Xu1998};\,\citen{Xie1998}), and a Bayesian approach (\citen{Yi2002}) and have been applied to a range of experimental datasets (\citen{Blanc2006};\,\citen{Chen2009};\,\citen{Rueckert2010};\,\citen{Coles2010};\,\citen{Steinhoff2011}).

In the field of evolutionary genetics, information from {QTL} analysis has been employed to infer the evolutionary forces acting on a particular trait. Here, the central question is whether natural selection acted on a trait during its evolutionary history. A more specific question is if the strength of selection is constant across a phylogeny, or whether it acted in a lineage-specific manner.  Several statistical tests make use of the data gained from {QTL} experiments to detect effects of natural selection. The test of Orr (\citen{Orr1998}) asks if the statistical distribution of alleles shows an excess of alleles that increase the value of the trait (``+'' alleles) in one line as a sign of selection. Orr's test was in turn assessed by Anderson and Slatkin (\citen{anderson2003}), finding that the test statistics is conservative, and Rice and Townsend~(\citen{RiceTownsend2012G3}), observing an unusual dependence of the test on the variance of the distribution of additive effects.
Based on Orr's approach, Fraser, Moses, and Schadt~(\citen{Fraser2010};\,\citen{Fraser2011}) used {QTL} statistics to detect a signal of non-neutral evolution in gene expression levels in different yeast strains.  The test of Rice and Townsend~(\citen{RiceTownsend2012}) combines {QTL} analysis with data from mutation accumulation experiments and asks if mutations seen between two lines tend to affect the trait more than those seen in experiments that accumulate largely neutral mutations.  However, currently no test uses the full statistical information from multiple-line {QTL} experiments.

In this paper we develop a statistical framework to test different evolutionary hypotheses for multiple {QTL} lines.  Using a systematic likelihood-based approach, we find that a multiple-line test has a higher statistical power to identify selection compared to the two line test. However, the consequences of multiple-line testing go beyond the mere increase of the number of observed loci. In two lines, the effect of lineage-specific selection turns out to be statistically indistinguishable from the bias introduced by testing traits with largest phenotypic differences from a pool of traits for selection. In three or more lines, evolutionary scenarios involving lineage-specific selection can generally be distinguished from such bias. We use this effect to search for lineage-specific selection in QTL data on different traits in species of the genus \textit{Mimulus} and in different maize lines. 

Our test follows the test of Orr (\citen{Orr1998}) in that we use a two state model at each locus and infer selection from the statistics of $+$ and $-$ alleles. Also we condition 
the allele statistics on the phenotypic difference to deal with a potential bias introduced by testing multiple traits. Unlike Orr, we use population genetic models to compare 
the empirical allele statistics with the statistics observed under different evolutionary scenarios. The approach of Rice and Townsend~(\citen{RiceTownsend2012}) is similar
in spirit, but uses information from mutation accumulation experiments, which go beyond standard QTL analysis. Numerical simulations performed to assess the statistical power of the test are similar to Rice and Townsend~(\citen{RiceTownsend2012G3}) (which however focus on the connection to the variance of the distribution of additive effects) and for the multiple testing simulations we use a scenario analogous to Anderson and Slatkin (\citen{anderson2003}).

In the following, we develop a log-likelihood score which quantifies the likelihood of neutral and selective hypotheses in an explicit evolutionary framework. We first explore our approach on artificial data and probe the efficiency of our test in the presence of different confounding factors. We then discuss the bias that trait selection can introduce into the 
allele statistics, and how in three lines or more the effects of natural selection can be distinguished from bias due to trait selection. 
Finally, we apply our test to floral quantitative traits in different \textit{Mimulus} species and to photoperiod traits in maize, finding evidence for lineage-specific selection that is 
not detectable in two lines.  \\

{\centering
\section{\large An $n$-line selection model}
}

In this section we construct a simple population genetic model of quantitative trait loci evolving in $n$ haploid populations in the weak-mutation regime with full recombination. 
Trait and fitness are a linear functions of the states of the  loci. The effects of inter-locus epistasis, simultaneous polymorphism and lack of recombination will be examined in 
section~\textit{Epistasis and multiple segregating loci}.  

Central to our analysis is a quantitative trait $T$ affected by $L$ loci labelled $l=1,\ldots,L$. Each locus is characterized by a genotype, and the genotype at each locus affects the trait in a particular way. As an example, consider a trait affected by a transcription factor. In that case, the regulatory region of a gene may be a sequence locus affecting the trait.  We can approximate the relationship between trait and locus by a two-state variable $q$ with states ``on'' (functional binding site in the regulatory region) or ``off'' (non-functional site).  We describe each locus $l$ by such an effect state (state for short) $q_l$. The effect state depends on the genotype and describes the effect a particular genotype at a locus has on the trait. In general, different genotypes at a locus correspond to the same state (there are many different sequences with a functioning binding site, and even more without). We denote the number of genotypes of a locus corresponding to state $q$ by $\omega_q$.

Due to the limitations of {QTL} mapping, the information on the effect state of a locus is indirect; in most cases 
it is not known what feature of the genotype determines the state of the locus. Instead, for each allele at a locus, {QTL} analysis gives the 
effect a particular allele has on the trait averaged over many crosses. In {QTL} studies using crosses between $4$ different lines~(\citen{Blanc2006};\,\citen{Coles2010}), most loci show a clear separation between alleles; the different alleles either decrease or increase the trait by a certain amount. For this reason, we restrict ourselves to a two-state model of loci, $q_{l}=\pm 1$, effectively focussing on the feature of a locus' genotype with the largest effect on the trait. An extension to more states is easily possible and may be required when analyzing a large number of lines: In a study using crosses between $25$ lines, loci harbouring alleles with several different effects on the trait have been observed~(\citen{Buckler2009}).
 
We assume a linear trait model (character model) without trait epistasis (inter-locus epistasis); the state at each locus contributes additively to the trait 
\begin{equation}
\label{eq:trait}
T(\{q_l\}) = \sum_{l=1}^L a_l q_l \ ,
\end{equation}
where the additive QTL effect $a_l$ specifies the contribution of locus $l$ to the trait. Without loss of generality we take $a_l\geq 0$, so $q_l=+1$ (termed the $+$-state) results in a higher trait value than $q_l=-1$ (the $-$-state). The additive effect $a_l$ of a locus is taken from experiments on multiple crosses between different lines, as is the state $q_l$ of a particular allele. $\{q_l\}$ denotes the set of effect states at all $L$ loci. 
We assume a linear Malthusian fitness (log-fitness) landscape 
\begin{equation}
\label{eq:fitness}
F(\{q_l\}) =  s T(\{q_l\}) =  s \sum_{l=1}^L a_l q_l
\end{equation}
with selection strength $s>0$, resulting in a selection coefficient $\sigma_l = N s a_l$ for each locus proportional to the additive effect $a_l$. Under this assumption, the effect of the state of a locus on both trait and fitness is independent of the states of other loci. This assumption will be examined and relaxed below. There is no environmental component and (for a diploid population) no dominance. 

We consider a simple population-genetic model describing a haploid population of effective population size $N$ in the weak-mutation regime with full recombination. 
In this regime, mutations appear at some rate, and are eventually either fixed or excised from the population. The arrival and fixation of mutations is a stochastic process, whose rate $\mu N (1-e^{-2 \Delta F})/ (1-e^{-2 N \Delta F})$ depends on the
fitness difference $\Delta F$ of the new allele relative to the pre-existing allele, the effective population size $N$ and the mutation rate $\mu$~(\citen{Wright1931};\,\citen{Kimura1962}).

At low mutation rates, most loci are monomorphic at a given point in time, but may differ between lines (due to mutations that fix in a given population before the next mutation occurs). The statistics of states $P(q)$ of a locus describes the probability that this locus in a given line is in state $q$. In the limit of long evolutionary times between lines, this statistics no longer changes with time, so the probability $P(q)$ is stationary (equilibrium).
Under neutral evolution, the equilibrium probability $P(q)$ depends only on the number $\omega_q$ of sequence variants of the locus corresponding to state $q$, $P(q)=\omega_q/(\omega_+ + \omega_-) = \frac{\exp\{\entropy q\}}{\exp\{\entropy\}+\exp\{-\entropy\}}$. The shorthand $\entropy = (1/2) \log(\omega_+/\omega_-)$ is called the multiplicity parameter of a particular locus.
In our example with the transcription factor binding site, the number of sequences with a functioning binding site $\omega_+$ is much lower than the number of sequences without 
such a site $\omega_-$, leading to $P(q=+1) \ll 1$ in the absence of selection. The multiplicity parameter of a locus quantifies the
asymmetry between $+$ and $-$ state in the absence of selection, and correspondingly the relative number of mutations at a locus increasing or decreasing the trait. 
Under selection, however, the equilibrium state statistics $P(q)$ depends also on the fitness difference between the two states and is given by (\citen{Iwasa1988};\,\citen{Berg2004};\,\citen{Sella2005})
\begin{equation}
P(q|N s a+\entropy)=\frac{e^{N s a q+\entropy q}}{e^{N s a+\entropy}+e^{-N s a - \entropy}} \ .
\label{eq:KimuraOhta}
\end{equation}
This result is valid in the low-mutation regime, but can be generalized~(\citen{Iwasa1988};\,\citen{Barton2009};\,\citen{Nourmohammad2013b}). A brief derivation is given in appendix II. The key assumption behind this result is that after long times since the last common ancestor a stationary distribution $P(q)$ is reached. This assumption will be examined in section \textit{Testing for selection at different evolutionary times}. In appendix I we derive results valid in the complementary regimes of short times since the last common ancestor.

For $n$ lines labelled $i=1,\ldots,n$, the joint probability distribution in the limit of long evolutionary time factorizes over lines, so the statistics of states for a given locus is
\begin{equation}
P(q_1,\ldots ,q_n|N s_1 a+\entropy,\ldots) = \frac{1}{Z} e^{\sum_{i=1}^n (N s_i a + \entropy) q_i}\,,
\label{eq:nlinesselective}
\end{equation}
where $Z= \sum'_{q_1,q_2,\ldots ,q_n=\pm1} e^{\sum_{i=1}^n (N s_i a + \entropy) q_i}$. Here, we need to consider one subtlety arising from {QTL} analysis based on crosses between individuals from different lines: In crosses, only the effects of loci differing in their state $q$ in at least two lines can be determined. For this reason, the two configurations $q_1=q_2=\ldots=q_n=\pm1$ remain unobservable. Thus, the sum in the normalizing factor $Z$ is over all states of the $n$ lines $q_1,q_2,\ldots,q_n$ excluding the cases $q_1=q_2=\ldots=q_n$ (indicated by $\sum'$).

Under the linear fitness model~(\ref{eq:fitness}), states at different loci are statistically independent, so the statistics of states over several loci is the product of (\ref{eq:nlinesselective}) over loci
\begin{equation}
P(\{q_{i,l}\}|\{Ns_i\},\{a_l\},\{\entropy_l\}) = \prod_{l=1}^{L_\text{div}}P(q_{1,l},\ldots ,q_{n,l}|N s_1 a_l+\entropy_l,\ldots) \ ,
\label{eq:nlinesselectiveall}
\end{equation}
where the number of loci with different states in at least two lines is denoted by $L_\text{div}$. The statistics of states at different loci may differ from this 
simple form for several reasons. The first is genetic linkage: here we assume free recombination between loci, as is standard in quantitative genetics (see below for an example with full linkage). A second reason is epistasis, which will be discussed in the section on \textit{Epistasis and multiple segregating loci}.  \\

{\centering
\section{\large Inference and hypothesis testing for different evolutionary scenarios}
}

The statistics of states~(\ref{eq:nlinesselectiveall}) can be used to infer the parameters of this model (selection strengths $N s_i$ for different lines and the multiplicity parameters $\entropy_l$ at different loci) from experimental data on the states $\{q_{i,l}\}$ across lines and loci and on the additive effects $\{a_l\}$. Denoting the position of the maximum $f(x^*)$ of a function $f(x)$ over $x$ by $x^*=\argmax\limits_x f(x)$, the maximum-likelihood estimates of the free parameters $\{Ns_i, \entropy_l\}$ are obtained by maximizing (\ref{eq:nlinesselectiveall}) with respect to the free parameters
\begin{equation}
\{Ns_i^*, \entropy_l^*\}=\argmax\limits_{\{Ns_i,\entropy_l\}} P(\{q_{i,l}\}|\{Ns_i\},\{a_l\},\{\entropy_l\}) \ .
\label{eq:maxlikeli}
\end{equation}

There are two limitations to the inference of multiplicity parameters and selection strengths. The first is that the number of lines $n$ limits in particular the inference of multiplicity 
parameters. For $n=2$ lines, the only observable loci are in states $(q_1,q_2)=(+-)$ or $(-+)$, so $P(q_1,q_2| N s_1 a+\entropy,\ldots)= e^{(s_1-s_2) a (q_1 - q_2)}/Z$. Hence the statistics of states does not depend on the multiplicity parameters, making their inference impossible. For $n>2$, the statistics of states depends on the multiplicity parameters, and the estimate of these parameters improves with increasing number of lines, since the size of the data  (number of loci times the number of lines) increases relative to the number of multiplicity parameters (one per locus). 
Second, selection strengths can only be determined relative to each other: The likelihood~(\ref{eq:nlinesselectiveall}) depends on the states $q_{l,a}$ via $\sum_{a,l} (N a_l s_a+\entropy_l) q_{a,l}$. Increasing all selection strengths uniformly by some $\bar{s}$ and decreasing each multiplicity parameter $\entropy_l$ by $N a_l \bar{s}$ thus leaves the likelihood unchanged. As a result, e.g. a situation where  selection strength $\bar{s}$ 
is uniform over the lines and multiplicity parameters are all zero is statistically indistinguishable from a multiplicity parameter  $\entropy_l=N a_l \bar{s}$ and neutral evolution. In the following 
we will focus on lineage-specific selection, and determine selection strengths relative to each other. Using further information on multiplicity parameters (for instance from mutation 
accumulation experiments~(\citen{RiceTownsend2012})), or further assumptions (for instance that multiplicity parameters be uncorrelated with effect sizes, or be on average non-negative) one can also obtain information on absolute selection strengths from~(\ref{eq:nlinesselectiveall}). 

When only few loci for a trait are known, the inference of all parameters may be unreliable due to overfitting. In this case it is convenient to restrict the parameter space  and test specific hypotheses against each other. For example, one can compare a scenario with uniform selection strength on all lines ($s_1=s_2=\ldots=s_n=s$) with a lineage-specific selection pattern ($s_1\neq s_2=\ldots=s_n=s$). The log-likelihood score 
\begin{equation}
S_{Q,P} = \sum\limits_{l=1}^{L_\text{div}} \ln\left( \frac{Q(q_{1,l},q_{2,l},\ldots ,q_{n,l}|N s_1^{*} a_l+\entropy_l^{*},\ldots)}{P(q_{1,l},q_{2,l},\ldots ,q_{n,l}|N s_1^{*'} a_l+\entropy_l^{*'},\ldots)}\right)\,,
\label{eq:LogScore}
\end{equation}
quantifies the evidence for two such evolutionary scenarios $P$ and $Q$ relative to one another. Both these scenarios are described by statistics of the form~(\ref{eq:nlinesselectiveall}) but differ in their parameter values. The score (\ref{eq:LogScore}) is positive if the distribution of states observed in a particular data set is more in agreement with the statistics of states in scenario $Q$ than in scenario $P$. For both these scenarios, the remaining selection parameters are estimated together with the multiplicity parameters according to eq.~(\ref{eq:maxlikeli}).

When two scenarios with different numbers of free parameters are tested against each other, the log-likelihood score is generally biased towards the scenario with more parameters. A simple way to correct this bias is the Bayesian information criterion (BIC) (\citen{Schwarz1978}). Under the BIC correction, the score (\ref{eq:LogScore}) is decreased by an offset $k/2 \ln L_\text{div} $, where $k$ is the excess number of parameters in model $Q$.\\

{\centering
\section{\large Increased statistical power in more than two lines}
}

There is a simple reason why the power of the selection tests increases when more lines are used. Since only loci with different states in at least two lines can be observed, a certain fraction of loci affecting the trait remain hidden from the analysis. For two lines, loci with the states $(q_1,q_2)=(++)$ and $(--)$ cannot be observed. For three lines there are only $2$ unobserved out of $8$ possible configurations and the fraction of unobserved loci decreases further with the number of lines. In general, the probability of a locus to remain unobserved in $n$ lines is given by $\gamma(n|s_i,\entropy) = \prod_{i=1}^n P(+1|Ns_i a + \entropy) + \prod_{i=1}^n P(-1|Ns_i a + \entropy)$, where the statistics of states $P(q|Ns_i a + \entropy)$ is given by~(\ref{eq:KimuraOhta}).

To probe the log-likelihood score \eqref{eq:LogScore} for a varying number of lines, we test selective and neutral hypotheses against each other on artificial data. For $n=2\ldots 6$ lines and $L=20$ loci, additive effects $\{a_l\}$ are drawn randomly from a gamma distribution (\citen{Orr1998};\,\citen{Zeng1992}). After choosing the effects $\{a_l\}$, their values are fixed and are taken to be known explicitly (in practice obtained through experiments using {QTL} crosses). Then we generate artificial QTL data under different scenarios, which we label for easy reference. In the first, neutral scenario $P_0$, the selection strength on all lines is zero ($s_1=s_2=\ldots =s_n=0$). In the second scenario $Q_1$, only line $1$ is under selection ($s_1= s$, $s_2=\ldots  = 0$). 

In each run, a set of states $\{q_{1,l},\ldots,q_{n,l}\}$ is drawn from the probability distribution~(\ref{eq:nlinesselective}) with fixed values of $Ns_i$ for each line and $\entropy_l$ for each locus corresponding to scenario $Q_1$, see caption for details. For the subset of loci with different states in at least two lines the log-likelihood score $S_{Q_1,P_0}$ (\ref{eq:LogScore}) is computed. To gauge the statistical significance of a given value of this score, we also estimate the probability of reaching the same score or higher under the neutral scenario $P_0$. This $p$-value measures the rate of false positives (type I error rate) and is computed by performing a large number of runs under the scenario $P_0$ to see what fraction of them gives a score matching or exceeding $S$. To gauge how frequently a positive score occurs in favour of scenario $Q_1$  with selection on line 1, 2 or 3, the configurations drawn from the null model $P_0$ are sorted according to their trait values $T_1,T_2,T_3$. 

\begin{figure}[htbp]
\begin{center}
\includegraphics[width=0.6\textwidth]{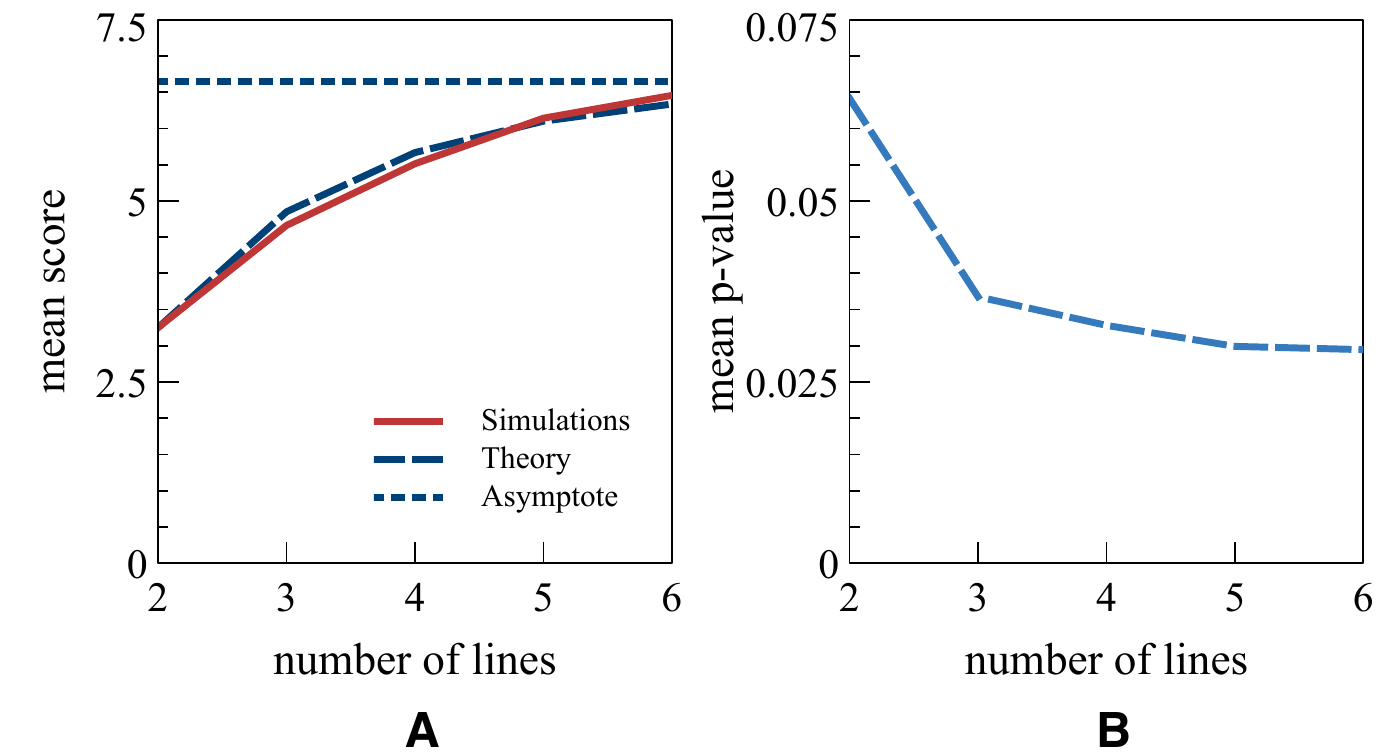}
\caption{\small {\bf Log-likelihood score (\ref{eq:LogScore}) and its statistical significance for a different number of lines.} (A) The expected log-likelihood score (\ref{eq:LogScore}) is shown for different numbers of lines at a fixed total number of $L=20$ loci. With an increasing number of lines on average fewer loci have the same state in all lines. Hence the number of detectable loci and thus the score increases with the number of lines. (B) The expected $p$-value for scenario $Q_1$ decreases with the number of lines. Parameters: average selection coefficient $\sigma=Nsa=1$ with mean additive effect $a=0.1$ per locus, $L=20$, multiplicity parameter $\entropy_l=\pm 0.2$ (each for half of the loci, respectively). Parameters of the gamma distribution of additive effects are $\alpha=2$, $\beta=20$.}
\label{fig:NumberOfLines}
\end{center}
\end{figure}

As expected, the log-likelihood score for the selective model $Q_1$ increases with the number of lines, while the mean $p$-value decreases, see Figure~\ref{fig:NumberOfLines}. This increase in statistical power due to the increased number of loci is a simple quantitative effect arising from an increase in the number of loci $L_\text{div}$ with different states in at least two lines.  The dependence of the score $S$ on the number of lines $n$ is approximately given by $S(n)=S_2 \frac{1-\gamma(n|s_i,\entropy)}{1-\gamma(2|s_i,\entropy)}$, where $S_2$ is the score for two lines. Since we average over many loci to obtain $S$, the multiplicity parameter $\entropy$ appearing in $\gamma (n|s_i,\entropy)$ has to be understood as an average multiplicity parameter over the loci. $S(n)$ is an increasing function of $n$, with the largest increase in score between $2$ and $3$ lines (see Figure~\ref{fig:NumberOfLines} A). The value of $S(n)$ saturates for large $n$ as all loci become detectable; the exact saturation value is $S_2\frac{1}{1-\gamma(2|s_i,\entropy)}$. Nevertheless, the number of detectable QTL, and hence the statistical signal of selection can remain small, even when the number of lines is large, if selection strength is so high in all $n$ lines that all or nearly all QTL have the same state in all lines. This can be seen from the expression for the fraction of unobserved loci $\gamma(n|s_i,\entropy)$, which tends to $1$ as all $s_i$ go to $\pm \infty$. If $N s_i g >2$ in all lines (which is the important quantity here), less than $5\%$ of the loci are observable (assuming $\entropy=0$). In practice, this particular problem can be remedied by including in the analysis one line with small selection pressure on the trait.  

While more lines bring more information, they also increase the experimental effort required to perform pairwise crosses between them. For this reason we also compare two- and three-line tests while keeping the total number of crosses constant. Given a fixed number of crosses that can be performed, should those crosses be concentrated on two lines, or should 
pairwise crosses of three lines be performed (with fewer crosses between each pair of lines)?

To compare two- and three-line tests on QTL mapping data at a fixed total number of crosses, we simulate a QTL model for three lines with $L_{div}=20$ loci differing in state between these lines under scenario $Q_1$. 100 single nucleotide polymorphism (SNP) markers are simulated with every fifth marker being linked to a QTL whose additive effect is drawn from a gamma distribution. We perform $M_{tot}$ crosses between line 1 and 2 for the two-line mapping and $M_{tot}/3$ crosses for lines 1 and 2, 1 and 3, and 2 and 3, respectively, for the three line mapping. The recombination probability between two adjacent markers is set to $0.25$ such that the QTL segregate mostly independently. We used the random forest mapping method as described in (\citen{Beyer2010}) to infer QTL positions and additive effects. We then used the QTL found by the mapping algorithm for our selection test comparing the selective scenario $Q_1$ against neutral scenario $P_0$.

\begin{figure}[htbp]
\begin{center}
\includegraphics[width=0.85\textwidth]{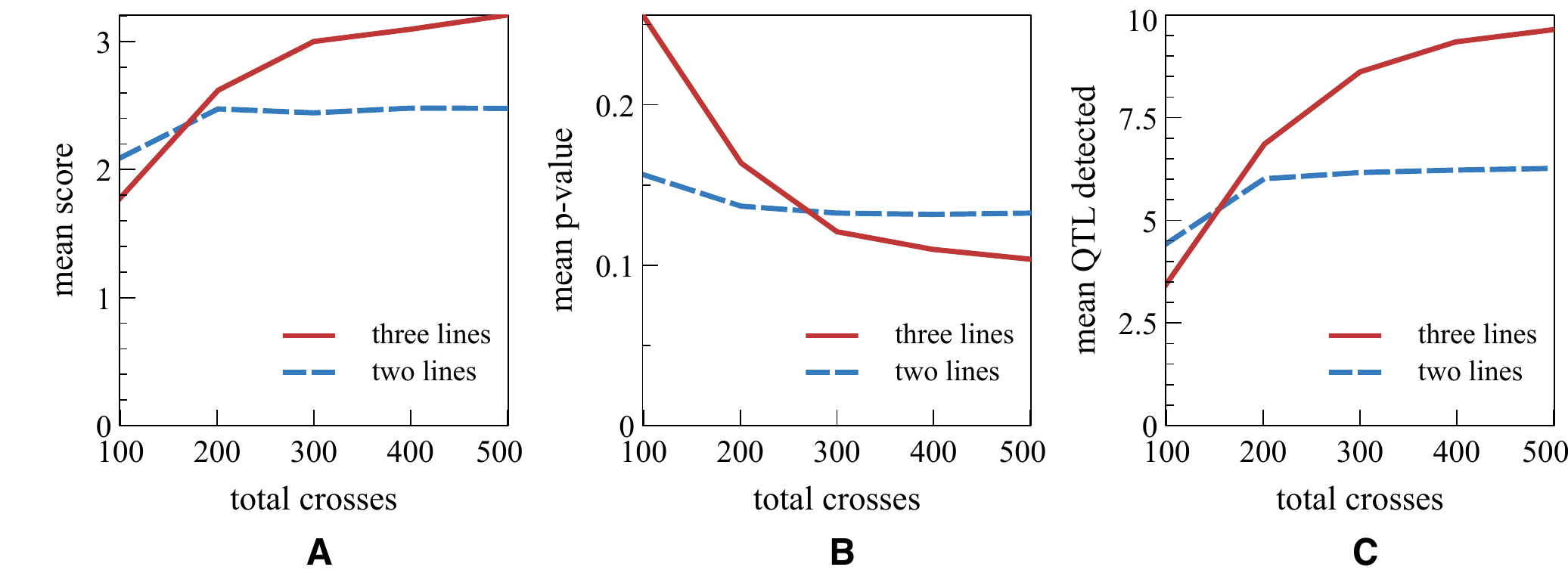}
\caption{\small {\bf Comparing two- and three-line test on QTL mapping data at a constant number of crosses.} (A) We plot the log-likelihood score (\ref{eq:LogScore}) averaged over 1000 runs with pairwise crosses either between three lines or between two lines against the total number of crosses. At around 200 crosses the three-line design is more effective and leads to higher scores. (B) The expected $p$-value for scenario $Q_1$ decreases faster with the number of crosses for three lines. (C) For two lines, fewer QTL can be detected since there are fewer diverged QTL available. Thus, for a high number of crosses (where all diverged QTL can be detected), more QTL can be detected with three-line design.  Parameters: $L_{div}=20$ loci diverged between three lines, 100 SNP markers, every fifth marker being a QTL with additive effects drawn from gamma distribution with $\alpha=10$, $\beta=10$. Selective scenario $Q_1$ was used with average selection coefficient $\sigma=Nsa=1$ with mean additive effect $a=0.1$ per locus.}
\label{fig:QTLmapping}
\end{center}
\end{figure}

The results in Figure~\ref{fig:QTLmapping} show that the three-line design is more effective at detecting selection given a sufficient number of crosses. However, for a small number of crosses the two-line test is more effective. The existence 
of two regimes can be understood as follows: At a large number of crosses between two lines, all or nearly all QTL that differ between these lines have been detected and further crosses do not yield new QTL. Between three lines, however, the number of diverged QTL is larger, so crosses between three lines can yield more QTL. The effect thus arises from the competition between detecting more QTL among those diverged between two lines, and having more diverged QTL available in three lines, but fewer 
crosses per pair of lines. In our simulations, the crossover between the two regimes lies around $M_{tot}=200-300$ crosses, which is a realistic number in QTL experiments, but of course this depends on details of the QTL mapping algorithm and simulation parameters.

So far, the increase in statistical power in multiple-line tests is due to an increase in the number of diverged loci $L_\text{div}$ with the number of lines. In order to address other, qualitative effects arising when the number of lines is increased, $L_\text{div}$ is kept fixed for the remainder of this paper.\\

{\centering
\section{\large Detection of selection}
}

\begin{figure*}[htbp]
\begin{center}
\includegraphics[width=0.95\textwidth]{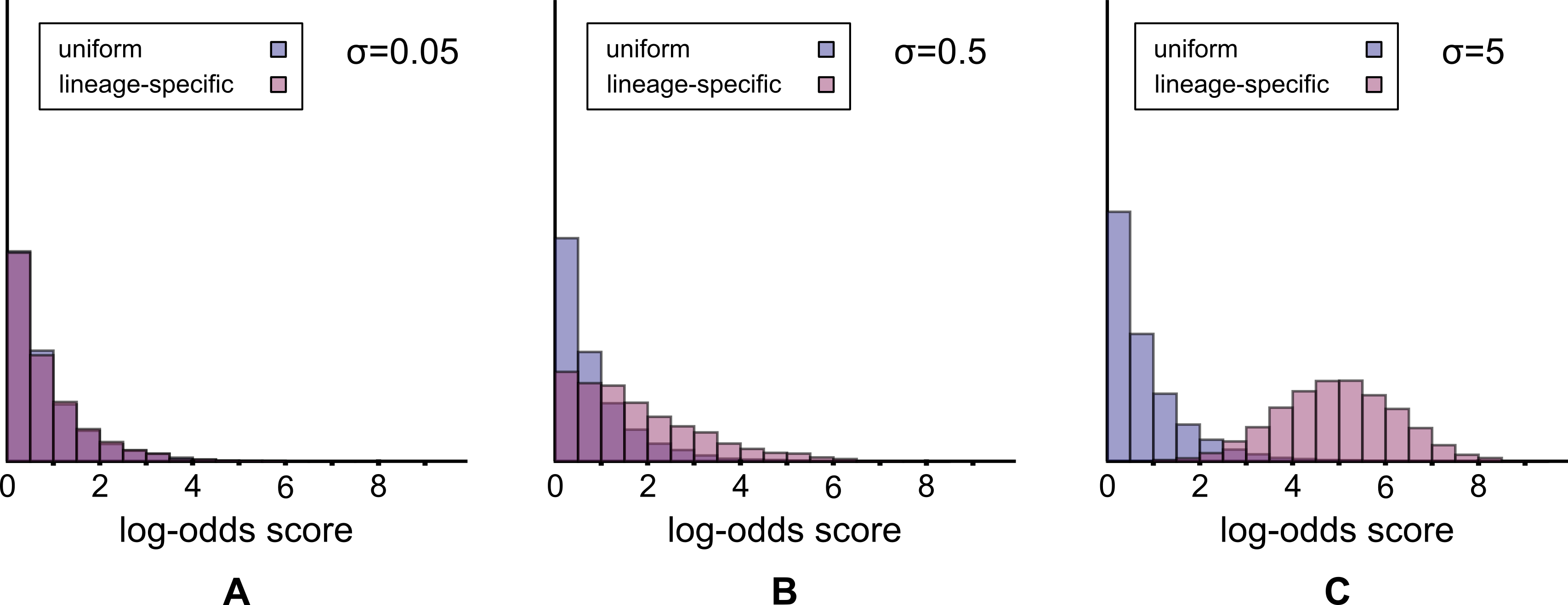}
\caption{\small {\bf Score statistics for a trait in two different evolutionary scenarios.} The distribution of the log-likelihood score (\ref{eq:LogScore}) under the neutral scenario $P_0$ and the lineage-specific selection scenario $Q_1$ are compared for different mean values of the selection coefficient $\sigma=Nsa$ (averaged over loci). (A) For small selection coefficients ($\sigma=0.05$), the score distributions (\ref{eq:LogScore}) under either scenario are nearly identical. (B and C) As the selection strength $s$ increases, the score distributions clearly separate. Parameters: three lines, $L_\text{div}=10$, $\sigma=Nsa=0.05$, $0.5$ and $5$, additive effects $\{a_l\}$ drawn from a gamma distribution with parameters $\alpha=2$ and $\beta=20$ (mean effect $a=0.1$ per locus), multiplicity parameter $\entropy_l=\pm 0.2$ each for half of the loci, respectively.}
\label{fig:OddsScore}
\end{center}
\end{figure*}

\begin{figure}[htbp]
\begin{center}
\includegraphics[width=0.45\textwidth]{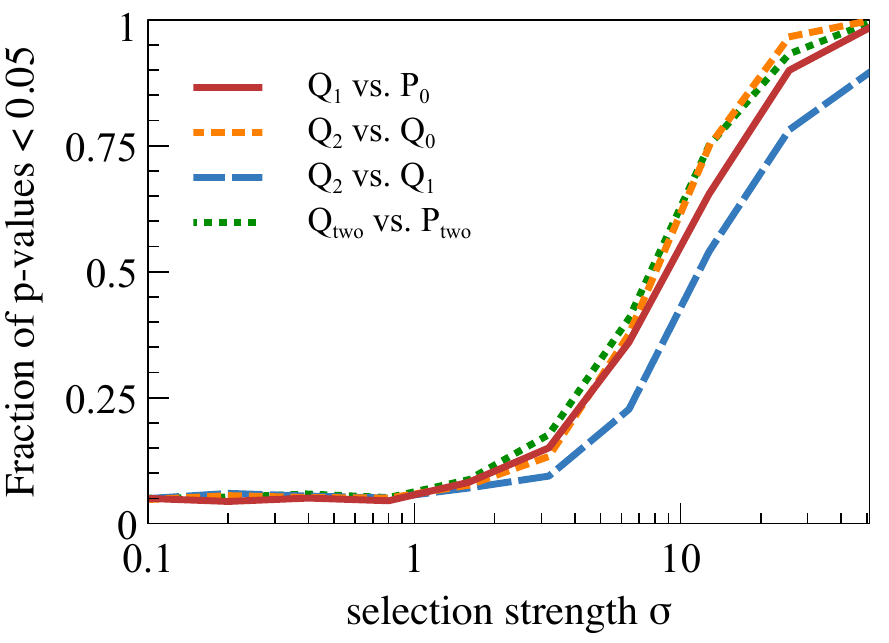}
\caption{\small {\bf Statistical significance of the tests at different levels of selection strength.} Selective scenarios are tested against the neutral hypothesis ($Q_1$ vs $P_0$, $Q_2$ vs $P_0$ and $Q_{\text{two}}$ vs $P_{\text{two}}$), as well as different selective scenarios against each other (lineage-specific scenarios $Q_2$ and $Q_1$). The fraction of instances where the log-likelihood score is statistically significant ($p<0.05$, see text) rises steeply with increasing selection strength (mean selection coefficient $\sigma=Nsa$ per locus). Parameters are as in Figure \ref{fig:OddsScore}.}
\label{fig:Simulations}
\end{center}
\end{figure}

To probe how well neutral and selective evolutionary scenarios can be distinguished we apply our test to artificial data generated under different hypotheses. The number of lines is set to $n=3$ and the number of diverged loci to $L_\text{div}=10$. In addition to the neutral scenario $P_0$ and the lineage-specific selection scenario $Q_1$ defined above, we also consider a scenario $Q_2$, where all three lines are under selection but in different directions ($s_1= +s/2$, $s_2=+s/2$, $s_3= -s/2$). We also test a selective two-line scenario $Q_{\text{two}}$ with a relative difference $\Delta s$ of selection strength between the lines against the neutral scenario $P_{\text{two}}$  ($\Delta s = 0$). Analogously to the previous section, scenarios $Q_1$ and $P_0$, $Q_2$ and $P_0$, $Q_2$ and $Q_1$, and $Q_{\text{two}}$ and $P_{\text{two}}$ are compared against each other. For the tests of scenario $Q_2$ against $Q_1$ the selection strengths are chosen to yield on average the same trait difference $\Delta T=(T_1-T_3)/2$.

Figure~\ref{fig:Simulations} shows that the log-likelihood score (\ref{eq:LogScore}) can clearly distinguish selective and neutral scenarios (see Figure~\ref{fig:OddsScore} and \ref{fig:Simulations}), as well as between different lineage-specific selection scenarios. As expected, the sensitivity of the test increases with selection strength. The test works in a reasonable parameter range, allowing one to infer selection strength with only few loci available ($L\gtrsim 4$ loci for $Ns a=1$) and a reasonable selection strength ($Ns a=1$ corresponds to a probability of $0.88$ for a locus to be in the $+$ state for $\entropy=0$).\\

{\centering
\section{\large Epistasis and multiple segregating loci}
}

The statistical framework comprising the equilibrium statistics of states~(\ref{eq:nlinesselectiveall}), the maximum-likelihood estimates of selection strengths~(\ref{eq:maxlikeli}), and the log-likelihood score~(\ref{eq:LogScore}) is built on a very simple population genetics model. In this section, we explore how the resulting test performs when specific assumptions behind this model are not fulfilled. To this end, we do finite-population simulations in a regime with multiple segregating loci, and look at two different kinds of epistasis between loci, phenotype and fitness epistasis. 

To model phenotype epistasis (character epistasis), we add a pairwise interaction term to the linear relationship between QTL states and the  trait~(\ref{eq:trait}), yielding 
\begin{equation}
\label{eq:traitepistasis}
T(\{q_l\})=\sum_{l=1}^L a_l q_l + \sum_{l,m=1}^L J_{lm} q_l q_m \ .
\end{equation}
$J$ is a $L \times L$ symmetric matrix describing the interactions between loci. The interaction coefficients $\{J_{lm}\}$ are drawn from the same gamma distribution as the effects $\{a_l\}$ (and are assigned random signs), however the average value of the $\{J_{lm}\}$ is varied relative to that of $a_l$ by multiplying them with a factor $J_0/L$. Then, for $J_0=1$ the cumulative contribution to the trait from the epistatic interaction $\sum_{l,m} |J_{lm}|$ is on average as large as the contributions from the linear term $\sum_l a_l$. The regime of large $J_0$ corresponds to significant epistasis: in this regime the trait value $T$ can change significantly with the change of state of a single locus. We assume that, as is generally the case, the epistatic interactions $\{J_{lm}\}$ are not known.

We perform numerical simulations using a Wright-Fisher model with and without trait epistasis. The Wright-Fisher model does not involve recombination, unlike the assumption of the selection test. Starting from a random initial configuration $\{q_l\}$ for $L=15$ loci, a Wright-Fisher model is simulated with three independent populations of 100 individuals each evolving over $M$ generations. At the end of each run, the configuration of loci with the largest fraction in the population is used to calculate the score (\ref{eq:LogScore}). We simulate both the selective scenario $Q_1$ and the neutral scenario $P_0$. We perform simulations both at high mutation rates leading to multiple segregating loci  (mutation rate $\mu=0.002$ over $M=3000$ generations, resulting in  $2\mu L N \ln N \approx 27.6 \gg 1$~(\citen{Wilke2004})) and in a second regime with low mutation rates ($\mu=2.5 \times 10^{-5}$ over $M=25000$ generations, with $2\mu L N \ln N \approx 0.35$), where there is typically at most a single segregating locus.

\begin{figure}[htbp]
\begin{center}
\includegraphics[width=0.45\textwidth]{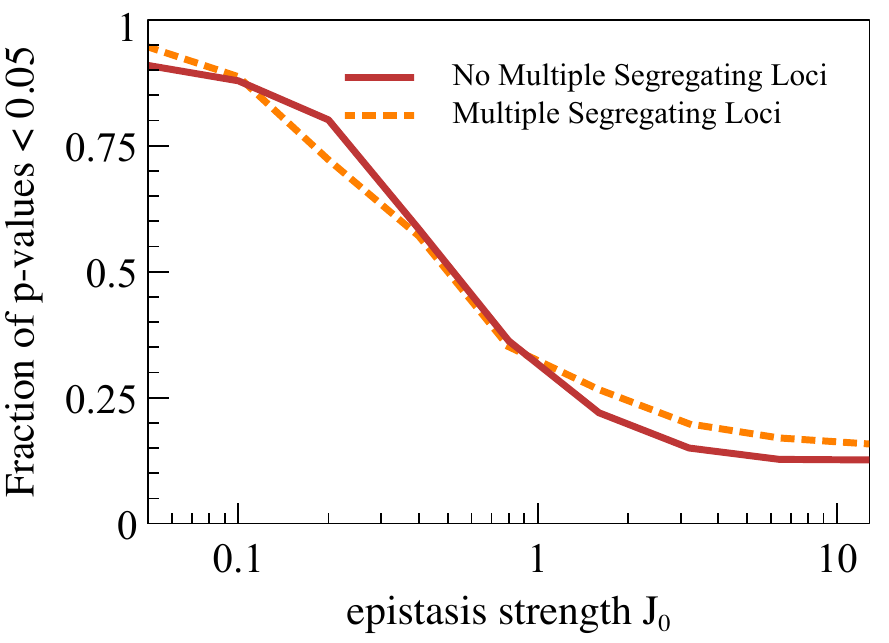}
\caption{\small {\bf The effect of phenotype epistasis.} We apply the selection test~(\ref{eq:LogScore}) to data generated under the model~(\ref{eq:traitepistasis}) with epistatic interactions between loci. With increasing epistasis strength $J_0$ the power of the test decreases, however for weak epistatic interactions $J_0 \ll 1$ the test retains most of its power. Simulation parameters: $n=3$ lines, $L=15$ loci, effective population size $N=100$, average selection coefficient $\sigma=Ns a = 10$, mutation rate $\mu=0.002$, $\mu=0.000025$ and $M=3000$, $M=25000$ generations in the case with and without multiple segregating loci, respectively. Other simulation parameters are as in Figure~\ref{fig:OddsScore}.}
\label{fig:PhenotypeEpistasis}
\end{center}
\end{figure}

As $J_0$ is increased, the effect of each locus on the trait becomes coupled to the states of other loci, and the linear trait model~(\ref{eq:trait}) becomes increasingly inaccurate. If the epistatic interactions in ~(\ref{eq:traitepistasis} were known, the trait model with epistasis~(\ref{eq:traitepistasis}) could be incorporated into the state statistics~(\ref{eq:nlinesselectiveall}) to restore the test's sensitivity.
As a result, the power of our test decreases with $J_0$, see Figure~\ref{fig:PhenotypeEpistasis}. Yet, for weak epistatic interactions $J_0 \ll 1$ the results of the test are only mildly affected. There is no significant difference in the power of the test between the regimes with and without multiple segregating loci in the regimes we examined.

\begin{figure}[htbp]
\begin{center}
\includegraphics[width=0.45\textwidth]{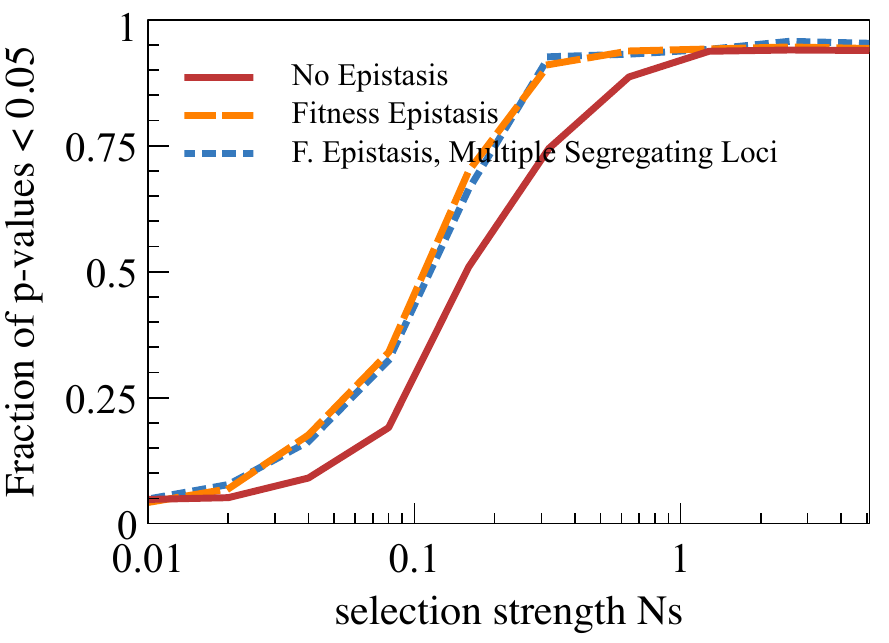}
\caption{\small {\bf The effect of fitness epistasis.} The selection test~(\ref{eq:LogScore}) is applied to data generated under a model with a quadratic fitness function, see text. The results are very similar to the result without epistatic interactions both in regimes with and without multiple segregating loci. Simulation parameters are as in Figure \ref{fig:PhenotypeEpistasis}.}
\label{fig:FitnessEpistasis}
\end{center}
\end{figure}

For fitness epistasis, we consider a quadratic fitness function $F=-s_e (T(\{q_l\})-T_0)^2$ in place of the linear function~(\ref{eq:fitness}). $T_0$ is the trait value giving maximal fitness and $s_e$ determines how quickly fitness decreases away from the maximum.
The fitness parameters $T_0$ and $s_e$ are chosen such that mean and variance of the distribution of trait values $T$ equal those  under the model~(\ref{eq:fitness}) without epistasis at a given value $Ns$. In this way, the scenarios with and without fitness epistasis can be compared directly. We again perform simulations in regimes with and without multiple segregating loci. Figure~\ref{fig:FitnessEpistasis} shows a very similar performance of the test on data generated under the linear and the quadratic fitness landscapes in both cases. 
This is because the test evaluates only on the probabilities of alleles at individual loci. Correlations between loci depend on the non-linearities of the fitness landscape (\citen{Nourmohammad2013b}), but they do not enter the test.

Beyond epistasis, the results of a QTL-based test for selection are potentially limited by pleiotropic effects: a subset of QTL of one trait may affect a second, unknown trait. If this unknown trait is under selection, but not the first, a QTL-based test may erroneously lead to the conclusion that the first trait is under selection (because some of its loci show a signal of selection induced by the second, unknown trait). Hence, the evidence for selection from QTL statistics pertains to the trait the loci were identified for, or some unknown trait with substantial overlap of QTL loci with the trait under study. Conversely, the trait under study may be under selection (favouring $+$ states, say), but some of its loci affect another trait also under selection, favouring $-$ states. If the second trait is unknown, the test would infer a selection strength on the first trait that is too low. With a small number of lines or loci, the signal of selection may even be lost altogether. \\

 {\centering
\section{\large Testing for selection at different evolutionary times}
}

Here, we probe the statistical power of the equilibrium test at different evolutionary times. The statistics of states~(\ref{eq:nlinesselectiveall}) was derived in the steady state and is reached a long time after the divergence of the different lines. This equilibration time depends, besides the mutation rate, on the strength of selection and the size of mutational targets. 
In a  regime of long evolutionary times  each locus has changed state many times since the last common ancestor. In a regime of short evolutionary times, most loci have not changed their state (and are thus not detected in crosses) and most diverged loci have undergone a single change of state over the phylogeny. With a sufficient number of lines, the two 
scenarios can be distinguished easily on the basis of the QTL states in all lines; in the limit of short times the states are compatible with a single mutation event in the phylogeny (for each diverged locus).

\begin{figure}[htbp]
\begin{center}
\includegraphics[width=0.37\textwidth]{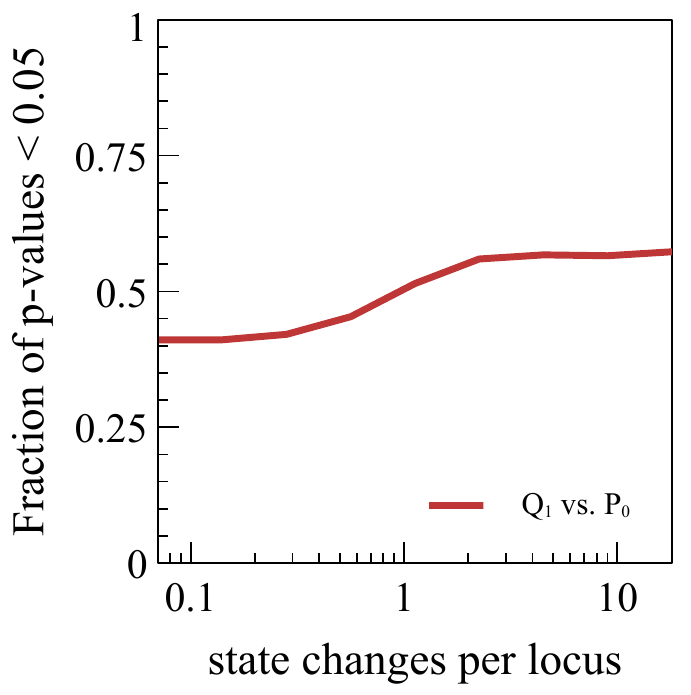}
\caption{\small {\bf The power of a test based on equilibrium statistics~(\ref{eq:nlinesselectiveall}) over different evolutionary times.} The significance of the three-line equilibrium selection test decreases only slightly with decreasing number of state changes per locus since the last common ancestor (corresponding to shorter evolutionary timescales). Both at intermediate times and even for short evolutionary times the equilibrium test retains most of its power. Parameters: evolutionary scenario $Q_1$ tested against $P_0$, average selection coefficient $\sigma=Nsa=1$, number of diverged loci $L_\text{div}=10$, $t_1=t_2=50$ steps, $\mu = 0.0002 - 0.06$.}
\label{fig:MidTimeRegime}
\end{center}
\end{figure}

We perform simulations analogous to the ones described in section \textit{Increased power in more than two lines}, but instead of drawing configurations $\{q_{1,l} , \ldots , q_{n,l} \}$ from the equilibrium distribution~(\ref{eq:nlinesselective}), we simulate for a number of $t$ time steps transitions between states at each locus with substitution rates  $\mu\frac{4Ns_i a}{1-e^{-4Ns_i a}}$ and $\mu\frac{-4Ns_i a}{1-e^{4Ns_i a}}$ for the transition from $-$ to $+$ and vice versa (\citen{Kimura1962}), see also appendix~I. The phylogenetic tree used for three lines is shown in Figure~\ref{fig:PhylogeneticTree}. To simulate the transition between short and long evolutionary times we vary the average number of substitutions per locus $\mu t$, 
but keep selection strength $Ns=10$ and the number of diverged loci $L_{\text{div}}=10$ fixed ($L_{\text{div}}$ is smaller than the total number of mutable {QTL} loci $L$ when the expected number of substitutions per locus is smaller than $1$). The score and p-value of the test~(\ref{eq:LogScore})
built on the assumption long evolutionary time is plotted against the average number of substitutions, see Figure~\ref{fig:MidTimeRegime}.
The statistical power decreases only slightly when going from long to short evolutionary times and the test retains some of its statistical power even as $\mu t$ goes to zero. The statistics of states in this limit of short evolutionary times is derived in appendix~I. \\

{\centering
\section{\large Multiple testing}
}

As emphasized by Orr (\citen{Orr1998}), a large trait difference between two lines alone is not sufficient evidence for lineage-specific selection. Often, traits in {QTL} experiments are picked from a larger pool of traits; among those, traits that diverged markedly between lines are chosen for further analysis, since this difference hints at lineage-specific selection. However, in a sufficiently large set of traits, neutral evolution alone would produce traits differing between lines. In such a trait, one would also observe an imbalance of states enhancing the trait value in one line and reducing it in the other. The bias in trait difference and the statistics of states resulting from a non-random choice from a set of traits is called ascertainment bias (\citen{Nielsen2003}). Ascertainment bias can lead to non-neutral evolution being attributed to a trait that evolved neutrally along with a set of other neutrally evolving traits. 

There are two ways to correct for this effect. If the total number of traits from which the observed trait is taken is known explicitly, one is faced with a standard 
multiple-testing problem. We look at this case first. However, if the trait is chosen from an ill-characterized set of traits, the situation is different. We follow the approach of Orr (\citen{Orr1998}) and consider the statistics of states conditioned on the observed trait difference. We will see that in this case there is a drastic difference between two and more than two lines.\\

{\centering
\subsection*{\large Holm--Bonferroni correction}
}

If the total number of observed traits is known, a standard multiple-testing correction can be applied. An example is gene expression levels, where traits are analyzed on a genome-wide level and the number of genes is known~(\citen{Fraser2010}). A suitable multiple-testing correction for this case is the Holm--Bonferroni correction~(\citen{Holm1979}), which has the advantage that no independence of the different hypotheses needs to be assumed. This is particularly important in {QTL} analysis, since different traits can be affected by the same genetic loci. The Holm--Bonferroni correction controls the familywise error rate (FWER), i.e. the false positive rate not only for a single trait but for a whole set of traits. If there are $m$ traits for which scenario $Q$ is tested against the null hypothesis of scenario $P$, one calculates the log-likelihood score $S_{Q,P}$ (\ref{eq:LogScore}) and the corresponding $p$-values $p_j$ for all $m$ traits. The traits are then ranked according to their $p$-values with the highest $p$-values first. Next, one searches for the first trait $j$ for which $p_j > \alpha / (m+1-j)$, where $\alpha$ is the significance threshold for the familywise error rate. 
Scenario $P$ can then be rejected for the traits $1,\ldots,j-1$, but not for traits $j,\ldots,m$.\\

{\centering
\subsection*{\large Conditioning on the trait difference}
}

Often, however, the size of the pool from which traits are picked is not known. Most traits from this pool remained unnoticed simply because they showed little difference between lines and were not recognized as interesting traits for investigation. The proposal of Orr (\citen{Orr1998}) for this case is to use, in place of~(\ref{eq:nlinesselectiveall}), a statistics of states conditioned on the empirical trait difference between two lines, that is to restrict the states to those giving rise to the observed $T_1-T_2$. In doing so, the part of the evidence for selection that comes from the trait difference between two lines is discarded. Orr writes the trait difference as $R=\sum_{l=1}^L a_l (q_{1,l}-q_{2,l})$ for the case of two lines. We generalize this notion to the case of $n$ lines and denote the maximal trait difference across two lines $R_{\text{max}}=\sum_{l=1}^L a_l (q_{1,l}-q_{2,l})$, where the lines are ordered such that line 1 has the largest trait value $T_1=\sum_{l=1}^L a_l q_{1,l}$ and line 2 has the smallest trait value $T_2$.

Our next step is to calculate the statistics of states conditioned on a particular value of $R_{\text{max}}$. This statistics can then be used in the log-likelihood score (\ref{eq:LogScore}) in place of the neutral null model. Our calculation is based on 
the principle of maximum entropy. This general principle applies to situations with incomplete knowledge on the probability distribution $p(x)$ of some variable $x$. This distribution must be consistent with any prior information on $x$ one might have (for instance the mean value of $x$), but otherwise it should be as unbiased as possible. The principle of maximum entropy posits that the distribution which best describes the incomplete state of knowledge 
 is the distribution which maximizes the information entropy $-\sum_x p(x) \ln p(x)$ with respect to $p(x)$, subject to the constraints resulting from prior information.
Stated in this form first by E. T. Jaynes (\citen{ETJaynes1957}), the principle of maximum entropy already appears at the core of statistical physics, where the distribution over configurations $x$ of a physical system are constrained by the mean energy $\langle E(x) \rangle = \sum_x E(x) p(x)$. The maximum entropy distribution in this case turns out to be the Boltzmann (exponential) distribution $p(x)\propto e^{-\beta E(x)}$, where $\beta$ is determined by the mean value of the energy $E(x)$. Other applications of the principle of maximum entropy are in image reconstruction (\citen{Narayan1986}), language modelling (\citen{Berger1996}), and neural networks (\citen{Mora2011}). In the context of quantitative traits, the principle of maximum entropy and the associated calculus of exponential distributions has been used to estimate unobserved allele frequencies and to infer selection from trait observables (\citen{PruegelBennett1994};\,\citen{PruegelBennett1997};\,\citen{Ruttray1995};\,\citen{Berg2004};\,\citen{Mustonen2005};\,\citen{Lassig2007};\,\citen{Mustonen2008};\,\citen{nBarton2009};\,\citen{deVladar2011};\,\citen{Nourmohammad2013a};\,\citen{Nourmohammad2013b}). Here, we use the principle of maximum entropy to derive the statistics of states conditioned on the largest trait difference $R_{\text{max}}$. A pedagogical example is given in the appendix.

Starting from the neutral null model $P_0$, we derive the neutral null model $P_{h}(q_1,\ldots,q_n|a,\entropy)$ conditioned on the trait difference, that is with an additional parameter $h$ determining the value of $R_{\text{max}}$. This distribution is obtained by maximizing the information entropy 
\begin{align}
H(P) = &-\sum'\limits_{q_i=\pm 1} P_{h}(q_1,\ldots,q_n|a,\entropy) \log \left(\frac{P_{h}(q_1,\ldots,q_n|a,\entropy)}{P(q_1,\ldots,q_n|\entropy)}\right) \nonumber\\
&+ \lambda_{0} \left(\sum'\limits_{q_i=\pm 1} P_{h}(q_1,\ldots,q_n|a,\entropy) - 1\right)\nonumber\\
&+ h \left( a \sum'\limits_{q_i=\pm 1} (q_1 - q_2) P_{h}(q_1,\ldots,q_n|a,\entropy) - \frac{R_{\text{max}}}{L_\text{div}} \right) 
\label{eq:threeline_information_entropy}
\end{align}
with respect to $P_{h}(q_1,\ldots,q_n|a,\entropy)$. Here, $P(q_1,\ldots,q_n|\entropy)$ refers to the neutral null model $P_0$. The sum over all possible states $q_i=\pm 1$, $i=1,\ldots n$ for a given locus again excludes the two unobserved states with $q_1=\ldots=q_n$. The maximisation is subject to two constraints, implemented by Lagrange multipliers; $\lambda_0$ to implement the normalization of $P_h$, and $h$ to implement the constraint that the largest trait difference $R_{\text{max}}$ equals the expected value $\sum_l \langle a_l (q_{1,l} - q_{2,i})\rangle$ under $P_h$, see appendix II. Setting the derivative of the information entropy (\ref{eq:threeline_information_entropy}) with respect to $P_{h}(q_1,\ldots,q_n|a,\entropy)$ equal to zero gives the state statistics of a locus with additive effect $a$ and multiplicity parameter $\entropy$  
\begin{equation}
P_{h}(q_1,\ldots,q_n|a,\entropy) = \frac{e^{h a (q_1 - q_2) + \entropy \sum_{i=1}^n q_i}}{ \sum'\limits_{\{q'_1,q'_2,\ldots ,q'_n=\pm1\}} e^{h a (q'_1 - q'_2) + \entropy \sum_{i=1}^n q'_i}}\,.
\label{eq:nlinesselectiveconditoned}
\end{equation}
The parameter $h$ is set such that the mean trait difference under (\ref{eq:nlinesselectiveconditoned}) (summed over all $L$ loci) equals the trait difference $R_{\text{max}}$ observed in the data. 

The maximum-entropy statistics $P_h$ conditioned on $R_{\text{max}}$ will be used to describe the statistics of states under neutral evolution and with ascertainment bias. The resulting log-likelihood score
\begin{equation}
S_{Q,P_h} = \sum\limits_{l=1}^L \ln\left( \frac{Q(q_{1,l},q_{2,l},\ldots ,q_{n,l}|a_l,\entropy_l)}{P_{h}(q_{1,l},q_{2,l},\ldots ,q_{n,l}|a_l,\entropy_l)}\right)
\label{eq:LogScoreh}
\end{equation}
compares evolution under selection and neutral evolution with ascertainment bias. This score 
depends on the ascertainment parameter $h$; extremizing the score with respect to $h$ sets the expected value of the trait 
difference under the conditioned model $P_{h}$ equal to the trait difference observed in the data. 

In the case of two lines it turns out that the probabilities for the two observable states $-+$ and $+-$, $P_{h}(q_1,q_2)= e^{ha(q_1-q_2)}/C$, are the same as for the selective model at equilibrium, $Q(q_1,q_2)= e^{\Delta s a (q_1 - q_2)}/C$ (the multiplicity parameters cancel for $q_1=-q_2$). Maximizing the score with respect to $h$, the statistics of states with ascertainment bias and under selection are exactly the same, making it impossible to distinguish selection from neutral dynamics and ascertainment bias~\footnote{A key difference of our log-likelihood score to Orr's test is that Orr not only uses the empirically observed additive effects $\{a_l\}$ available from crossing experiments, but also additive effects drawn from a  plausible distribution $P(a)$.  Orr's test can appear to yield significant results when calculating the trait difference $R$ using the additive effects empirically determined from crosses, but using a different set of additive effects drawn from some distribution $P(a)$ for $p$-value computations. Consistent with this, Rice and Townsend found that the outcome of Orr's test strongly depends on the assumptions made on that distribution and that the test can produce nonsensical results (\citen{RiceTownsend2012G3}) in particular cases. 
}. As a result, the log-likelihood score comparing evolution under selection at equilibrium with the neutral statistics conditioned on the observed trait value is exactly zero. Hence for two lines at equilibrium it is not possible to statistically distinguish neutral evolution 
with ascertainment bias from the effect of selection. 

\begin{figure}[htbp]
\begin{center}
\includegraphics[width=0.45\textwidth]{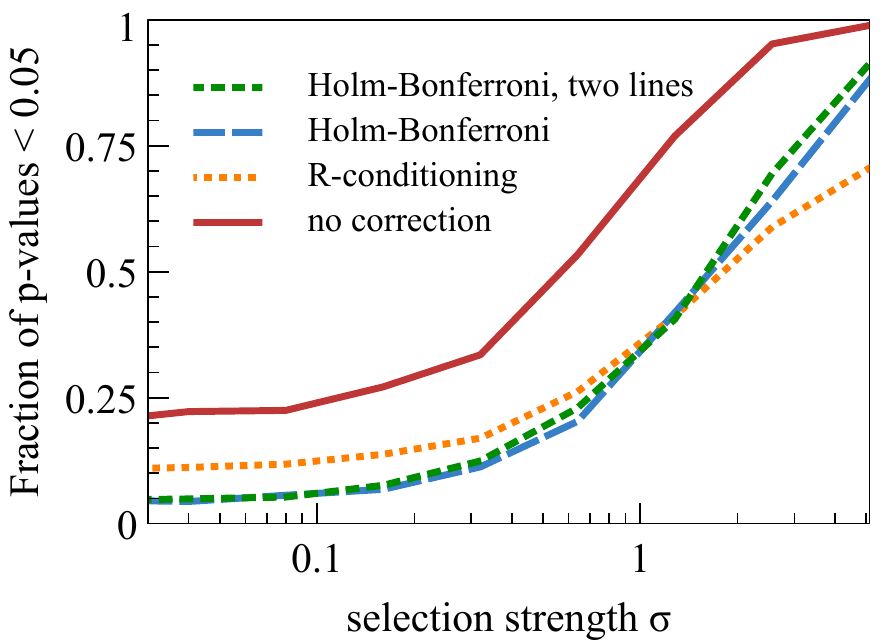}
\caption{\small {\bf Comparing tests with different multiple-testing corrections.} The statistical significance obtained without a multiple-testing correction, the Holm--Bonferroni correction, and by conditioning on $R_{\text{max}}$ are compared with each other. For three lines, the corrected tests both have a lower statistical significance (i.e. a higher p-value) but also a lower false positive rate (type I error rate). The false positive rate can be read off  at the very left of the plot as the fraction of significant outcomes under neutrality ($\sigma = 0$). The high false positive rate of $0.20$ without correction is reduced to $0.044$ for the Holm--Bonferroni correction and to $0.11$ for conditioning on $R_{\text{max}}$. Conditioning on $R_{\text{max}}$ gives a higher true positive rate for small selection strength than the Holm--Bonferroni correction but results in a higher false positive rate for a given significance threshold $\alpha$, see text. Parameters: $m=5$ traits, significance threshold $\alpha=0.05$ in all three cases. The other parameters are as in Figure \ref{fig:OddsScore}. }
\label{fig:SimulationsMultipleTesting}
\end{center}
\end{figure}

This situation is fundamentally different for more than two lines. For more than two lines the statistics of states in the selective scenario in equilibrium (\ref{eq:nlinesselectiveconditoned}) differs from the neutral scenario and the score (\ref{eq:LogScoreh})  generally gives non-zero results both at equilibrium and at short evolutionary times. (However, again there is a particular selection scenario, $s_1=+s$, $s_2=-s$, $s_3=0$, $\ldots,s_n=0$, that is not distinguishable from neutral evolution conditioned on $R_{max}$.)

To test these different approaches to the multiple-testing problem, we examine a multiple-testing scenario where a trait is picked from a larger set of traits. This multiple-testing scenario follows the lines of Anderson and Slatkin~(\citen{anderson2003}). First, states $\{q_{i,l}\}$ are drawn at random for $m=5$ traits and three lines evolving neutrally. Then the traits are sorted according to the maximal trait difference $R_{\text{max}}$ across lines. The trait with the highest $R_{\text{max}}$ is tested for selection using selective scenario $Q_1$ against the neutral scenario. We do this in three ways: using the score~(\ref{eq:LogScore}) without a multiple-testing correction, by applying the Holm--Bonferroni correction assuming the number of traits is known, and by conditioning on $R_{\text{max}}$ using (\ref{eq:LogScoreh}). Repeating this procedure many times over, we compute the false positive rate (type I error rate) for all three approaches, see Figure \ref{fig:SimulationsMultipleTesting}. Second, we generate the statistics of states of one trait under the selective scenario $Q_1$ and for the other traits under the neutral scenario $P_0$. Then, we determine how often the trait under selection is correctly identified by the different approaches (true positive rate) with a $p$-value less than $0.05$ ($0.05/m$ for Holm--Bonferroni). Figure~\ref{fig:SimulationsMultipleTesting} shows that, as expected, a test without correction yields the highest rate of true positives. Yet, it also suffers from the highest false positive rate, since many neutrally evolving traits happen to have a high $R_{\text{max}}$ leading to a high score~(\ref{eq:LogScore}). The Holm--Bonferroni method and the conditioning on $R_{\text{max}}$ both have lower false positive rates. This result for the conditioning on $R_{\text{max}}$ is in accord with Anderson and Slatkin~(\citen{anderson2003}), who found that the Orr test, which uses a similar correction scheme, also led to conservative test statistics. Since the false positive rate of the Holm--Bonferroni method is the lowest, it is to be preferred when the size of the pool of traits is known.

While the maximum trait difference $R_{\text{max}}$ is a plausible observable on the basis of which traits can be selected from a larger pool, it is by no means the only one. 
For instance, with three lines, traits could in principle be selected based on the difference between the trait in line $1$ and the trait mean in line $2$ and $3$, $R_{\Delta}=T_1- \frac{T_2 + T_3}{2}$. 
One would use this observable when looking specifically for traits with lineage-specific selection acting on line $1$. For $s_2=s_3\equiv s_0$ the fitness~(\ref{eq:fitness}) can be written
\begin{equation}
F(T_1, T_2, T_3)= \bar s (T_1+ T_2 + T_3) + \hat s \left ( T_1 - \frac{T_2+ T_3}{2} \right )
\end{equation} 
with  $\bar s = (s_1 + 2 s_0)/3$ and $\hat s = s_1 - \bar s = 2 (s_1 - s_0)/3$. The maximum entropy distribution conditioned on $R_{\Delta}$  is $\exp\{\lambda R_{\Delta}\}$ 
(up to a normalizing constant) and thus again differs from the equilibrium distribution $\propto \exp\{\beta F(T_1, T_2, T_3)\}$, except in the special case $\bar s=0$. \\

{\centering
\section{\large Selection on plant quantitative traits}
}

In this section, we apply the multiple-line selection test to data from two studies of plant quantitative traits. Our first example is based on {QTL} data on corolla (petal) sizes in three different plant species of the genus~\textit{Mimulus}. \textit{M. guttatus}, \textit{M. platycalyx}, \textit{M. micranthus} are labelled lines $1,2$ and $3$ respectively. At each locus detected by Chen (2009), it turns out there are two alleles with very similar effect on the trait (within experimental error), and one allele with a significantly different effect. If there is a single high allele, we assign it the $+$-state, while the two other alleles are assigned the $-$-state, and vice versa. Additive effects for the states are computed by averaging the additive effects listed for different alleles over alleles corresponding to the same state. The resulting states and additive effects for the corolla width and corolla length trait are listed in Table~1. 

Log-likelihood ratios for the pairwise comparisons of the evolutionary scenarios $P_0$, $P_h$ and $Q_1$ are calculated as described in Section \textit{Inference and hypothesis testing for different evolutionary scenarios}. These scenarios describe neutral evolution, neutral evolution in the presence of ascertainment bias, and lineage-specific selection, respectively. 
For each scenario, the multiplicity parameters and (in case of scenario $Q_1$) selection strength are calculated according to (\ref{eq:maxlikeli}).
Where applicable, we use the Bayesian information criterion described above to correct scores for different numbers of free parameters of the underlying models. This leaves $p$-values unaffected. When testing against a neutral scenario, we use either scenario $P_h$ (conditioning on $R$) or scenario $P_0$ (Holm--Bonferroni correction). In the first case, we condition the null model on the pair of lines with the highest trait difference for each trait. For the Holm--Bonferroni test we take the ad-hoc choice of $m=5$ as the total number of traits in this dataset since there are $5$ different traits analyzed in the {QTL} experiment in (\citen{Chen2009}). However, this choice is artificial as we do not know the potentially much larger set of traits these $5$ traits were chosen from.

\begin{table}[htbp]
\centering
 \setlength{\tabcolsep}{5pt}
  \begin{tabular}{cccc}
 additive effect $a_l$   & \textit{M. gutt.} & \textit{M. platy.} & \textit{M. micr.} \\
  \hline
  corolla width [mm]: & & & \\
  0.41 & $-$ & $+$ & $-$ \\
  0.74 & $+$ & $-$ & $-$ \\
  0.39 & $+$ & $-$ & $+$ \\
  0.59 & $+$ & $-$ &  $-$\\
  0.28 & $+$ & $+$ & $-$ \\
  0.64 & $+$ & $-$ & $-$ \\
  0.37 & $+$ & $-$ & $+$ \\
  \hline
  corolla length [mm]: & & & \\
  0.67 & $+$ &	$-$ & $-$\\
  0.41 & $+$	& $+$ & $-$\\
  0.21 & $+$	& $-$ & $+$\\
  0.60 & $+$	& $-$ & $-$\\
  0.27 & $-$	& $+$ & $+$\\
  0.51 & $+$	& $+$ & $-$
\end{tabular}
\label{tab:MimulusData}
\caption{\small Additive {QTL} effects for two flower traits of the Mimulus species \textit{M. guttatus}, \textit{M. platycalyx} and \textit{M. micranthus} estimated from (\citen{Chen2009}).}
\end{table}

We start with the corolla width trait, where $7$ {QTL} have been identified along with their additive effects (\citen{Chen2009}). Comparing scenario $Q_1$ against $P_h$ described by (\ref{eq:nlinesselectiveconditoned}) gives a log-likelihood score (\ref{eq:LogScoreh}) of $S_{Q_1,P_h}=0.38$ in favour of the selective scenario. We test the significance of this score by repeated simulations under scenario $P_h$ at fixed additive effects $\{a_l\}$. The ascertainment parameter $h$ is set such that the conditioned neutral model gives on average the trait difference $R_{12}$ observed in the data. For each configuration drawn from $P_h$ we sort the lines according to their trait values $T$. In this way we account for the possibility that under neutrality fluctuations create patterns of lineage-specific selection in any of the lines (rather than only in what is called line $1$ here). A $p$-value of $p=0.13$ is obtained. 

The unconditioned test together with the Holm--Bonferroni correction yields a similar result. Testing scenario $Q_1$ against $P_0$ the score $S_{Q_1,P_0}=1.90$ corresponding to a $p$-value of $p=0.05$ is obtained. With the Holm--Bonferroni correction, however a more stringent $p$-value cutoff $p<\alpha/m=0.01$ for the family-wise error $\alpha=0.05$ and $m=5$ has to be applied.

A preference for a selective model is in agreement with the different reproductive modes of these species (\citen{Chen2009}): line 1 reproduces predominantly by outcrossing (so that large floral characters are needed to attract pollinators), whereas line 2 and line 3 are mostly self-pollinating (but still maintain a certain degree of outcrossing). In the latter species, large petals are less indispensable for reproduction, but nevertheless require resources to develop and maintain. 

Next, we examine the corolla length trait, where $6$ {QTL} were observed (see Table~1) and the maximal trait difference is $R_{13}$ is between lines 1 and 3. Here, the comparison to the neutral null-model yields the score $S_{Q_1,P_h}=-0.43$ ($p=0.54$), so the neutral hypothesis cannot be rejected, similar for the Holm--Bonferroni procedure ($S_{Q_1,P_0}=1.04$, $p=0.14$ implying a substantial family-wise error). 

For comparison, we also apply Orr's sign test (\citen{Orr1998}) (not the equal effects version) to this dataset. Since the Orr-test is a two-line test, we apply it to the two lines with the largest trait difference, where one would expect the strongest signal for selection. Following Orr, the additive effects $\{a_l\}$ are taken from a gamma distribution whose parameters for each trait are estimated by maximum likelihood. Then the probability to find at least the observed number of $+$-states in the high line given the observed trait difference $R$ or greater is calculated according to eq. (4) in Orr's paper (\citen{Orr1998}). For the corolla width trait, 5 out of 6 diverged loci in line $1$ and $2$ have the $+$-state. Here, the Orr-test returns a $p$-value of $p=0.42$. For the comparison of line 1 and line 3 the test gives $p=0.29$. For the corolla length trait 4 out of 5 diverged loci are in the $+$ direction between line 1 and line 3 and 3 out of 4 diverged loci in the $+$ direction between lines 1 and 2. The Orr-test yields $p$-values of $p=0.48$ and $p=0.72$, respectively.

\begin{table}[htbp]
\centering
 \setlength{\tabcolsep}{5pt}
  \begin{tabular}{ccccc}
  additive effect $a_l$  & B73& B97 & CML254 & Ki14\\
  \hline
  GDDTA [GDD]: & & & &\\
  4.73 & $-$ & $-$ & $+$ & $+$\\
  3.85 & $-$ & $-$ & $+$ & $-$ \\
  4.43 & $-$ & $-$ & $+$ & $-$ \\
  11.13 & $-$ & $-$ &  $+$ & $+$\\
   \hline
  GDDTS [GDD]: & & & &\\
  6.33 & $-$ & $-$ & $+$ & $+$ \\
  6.20 & $-$	& $-$ & $+$ & $-$ \\
  4.68 & $+$	& $-$ & $+$ & $+$ \\
  5.68 & $-$	& $-$ & $+$ & $+$ \\
  \hline	
  plant height [cm]:& & & &\\
  1.10 & $-$ & $-$ & $+$ & $+$\\
  1.25 & $+$ & $+$ & $-$ &  $-$\\
  1.73 & $+$ & $-$ &  $+$ & $+$\\
  2.10 & $-$ & $+$ & $+$ & $-$\\
  \end{tabular}
\label{tab:MaizeData}
\caption{\small Additive {QTL} effects for three quantitative traits in the four maize lines B73, B97, CML254, and Ki14 estimated from (\citen{Coles2010}): growing degree day to anthesis (GDDTA), growing degree day to silking (GDDTS), and plant height.}
\end{table}

Our second example is based on {QTL} data on photoperiod response traits of four different maize strains. The photoperiod response of a trait is defined as the trait difference observed between specimens grown in an environment with long days and specimens grown in a short-day environment. We consider the traits `days to anthesis' (time from planting to full flower development) and `days to silking' (silk emergence in maize), both measured in growing degree days (daily average temperature above a threshold temperature of $10^{\,\circ}$C cumulated over days of growing). For comparison, we also look at plant height, which is not directly linked to day length. 
For maize it has been shown that the architecture of quantitative traits such as flowering time and leaf size accurately follows a model with additive trait effects and only weak epistatic effects (\citen{Buckler2009};\,\citen{Tian2011}).
In Coles et al. (2010), the additive effect of alleles from different {QTL} and the corresponding experimental errors are given. For each locus it is specified which lines harbour an allele with the same effect on the trait (within experimental error). 
As in \textit{Mimulus} above, most of the loci show alleles that have one of two experimentally distinguishable effects on the trait. In those cases, the states $+$ and $-$ can be unambiguously assigned to each line and locus, the resulting values for $\{q_l\}$ and $\{a_l\}$ are collected in Table~2. Yet, about one third of the loci show more than two significantly different effects on the trait, or have one line where the experimental error on the effect on the trait is so large that it cannot be assigned unambiguously to one of two states.
Loci with such unclear assignment of states are excluded from the analysis.

Two of the lines in (\citen{Coles2010}) (B73, B97) are taken from temperate climates featuring long days in summer and short days in winter, while the other two (CML254, Ki14) are taken from tropical environments with constant length of day over the year. Thus we use as the simplest evolutionary scenario $Q_4$ ($Ns_{\text{B73}}= -Ns$, $Ns_{\text{B97}}=-Ns$, $Ns_{\text{CML254}}= +Ns$, $Ns_{\text{Ki14}}= +Ns$), with only a single free parameter $s$. We compare this selective scenario $Q_4$ against the null model $P_{0}$ from (\ref{eq:nlinesselectiveall}) with $n=4$. 

We first consider the `growing degree day to anthesis' (GDDTA) trait, which measures the time to full flower development. For tropical lines, which are not adapted to long day length, the flowering time is reduced for specimens grown in temperate latitudes compared to tropical environments (\citen{Coles2010}). For the temperate lines no difference in flowering time is observed between the different environments. For this trait, $4$ out of $7$ loci show a clear two-state pattern. We first apply scenario $P_h$ conditioned on $R_{32}$. In this case, the straightforward maximum likelihood estimate of the parameter $h$ fails, since all states in the high line are $+$-states and all states in the low line are $-$-states, leading to a diverging $h \rightarrow \infty$. We use a lower-bound estimate for $h$ by determining the value $h$ for which the probability to see this extreme configuration equals $p_e$. $p_e=0.1$ is chosen to obtain a conservative estimate for $h$. For consistency, $Ns$ is determined in the same way. The log-likelihood score (\ref{eq:LogScore}) then gives $S_{Q_4,P_{0}}=2.77$ ($p=0.07$) in favour of the selective scenario. The Holm--Bonferroni correction yields a result consistent with this ($S_{Q_4,P_{0}}=5.06$, $p=0.045$).

For the `growing degree day to silking' (GDDTS) trait, with $4$ two-state loci out of $6$, the score $S_{Q_4,P_{h}}=2.77$ ($p=0.048$) favours the selective scenario over the neutral null model as well. Again $P_h$ is conditioned on $R_{32}$ and the lower bound for $h$ is used as described above. Using Holm--Bonferroni one obtains a similar result ($S_{Q_4,P_{0}}=5.06$, $p=0.030$). The `plant height' trait, on the other hand, with $4$ two-state loci out of $6$, yields a score $S_{Q_4,P_{0}}=-0.61$, $p=0.42$ under conditioning and $S_{Q_4,P_{0}}=-0.90$, $p=0.022$ with Holm--Bonferroni in favour of the neutral model. Here, $h$ was again determined by maximum likelihood and the conditioning was on $R_{34}$. The other traits investigated in the study (\citen{Coles2010}) (growing degree day anthesis-silking interval, ear height and total leaf number) have fewer two-state loci ($\leq 3$) and none of these traits show a significant support for either of the two hypotheses (data not shown). 

Again we also apply Orr's test for comparison. We compare the two lines B73 and CML254, which show the largest trait difference both in the GDDTA and the GDDTS trait. For the GDDTA trait, 6 out of 6 diverged loci have the $+$-state, giving a $p$-value $p=0.13$. For the GDDTS trait, 5 out of 5 diverged loci go in the $+$ direction with $p=0.2$. A summary of the results can be found in Table~3.

In both case studies, the statistical significance of the evidence for a particular evolutionary scenario is limited by the number of identified trait loci. With a higher number of crosses in the original studies, identifying more trait loci, we expect a stronger statistical signal.
\\

\begin{table}[htbp]
\centering
  \begin{tabular}{lccc}
  \textit{Mimulus} study & $Ns_1$ & $S$ & $p$  \\
  \hline
  corolla width & &  &\\
  $Q_1$ vs. $P_h$  & 2.2 &  0.38 & 0.13\\
  $Q_1$ vs. $P_0$  & 2.2 &  1.9 & 0.05\\
  corolla length & & &\\
  $Q_1$ vs. $P_h$  & 2.2 & -0.43 & 0.54\\
  $Q_1$ vs. $P_0$  & 2.2 & 1.0 & 0.14\\  
	& & &\\
  Maize study & $Ns_4$ &  $S$ & $p$\\
  \hline
  GDDTA & & &\\
  $Q_4$ vs. $P_h$  & 40 &  2.8 & 0.07\\
  $Q_4$ vs. $P_0$  & 40 &  5.1 & 0.045\\
  GDDTS & & &\\
  $Q_4$ vs. $P_h$  & 27 &   2.8 & 0.048\\
  $Q_4$ vs. $P_0$  & 27 &  5.1 & 0.030\\
  plant height & & &\\
  $Q_4$ vs. $P_h$  & 0.77 &  -0.61 & 0.42\\
  $Q_4$ vs. $P_0$  & 0.77 &  -0.90 & 0.022\\
  \end{tabular}
\label{tab:DataResultSummary}
\caption{\small Summary of results for the QTL data of the \textit{Mimulus}~(\citen{Chen2009}) and maize studies~(\citen{Coles2010}). Different evolutionary scenarios are tested against each other using both conditioning on the trait difference ($P_h$) as well as the Holm--Bonferroni correction ($P_0$). $Ns_1$ and $Ns_4$ denote the inferred selection strengths of the $Q$-scenarios, $S$ is the log-likelihood score obtained and $p$ the corresponding $p$-value. In \textit{Mimulus}, corolla width shows some evidence of selection, in maize the photoperiod response traits GDDTA and GDDTS.}
\end{table}

{\centering
\section{\large Conclusions}
}

In this paper, we developed a statistical framework to quantify the evidence for different evolutionary scenarios from {QTL} data for more than two lines. We find that using more than two lines not only increases the statistical power of selection tests, but also their scope: for more than two lines, signals of selection can be distinguished from the effects of ascertainment bias. We applied our test to {QTL} data on floral characters in different \textit{Mimulus} species and photoperiod response traits in maize.

Applying our test to very large numbers of lines poses interesting challenges in connection with the number of alleles per locus and the rapid growth of the number of possible evolutionary scenarios.  At the same time, the need for experimental crosses between three or more different lines is a major bottleneck of the multiple-line test. Due to the additional experimental work involved, there are currently few datasets on {QTL} and their additive effects in more lines than two. However recent studies employing crosses of 25 maize lines and detecting around 30-40 {QTL} per trait give a promising outlook to the future (\citen{Buckler2009};\,\citen{Tian2011}).

A possible application of this test is the inference of gene expression adaptation using expression {QTL} (e{QTL})~(\citen{Fraser2011}). Since the number of e{QTL} is typically small for a single gene, the test could be applied on gene modules, e.g. genes belonging to the same pathway or protein complex, allowing one to infer selection on individual pathways.
Another future perspective for this method may arise if genome-wide association studies (GWAS) with fully sequenced organisms enable the inference of causal mutations behind {QTL} effects (\citen{Manolio2009};\,\citen{Mackay2009}), allowing one to apply multiple-line tests without the need to perform crosses between different lines~(\citen{Fraser2013}).\\

\section*{\large Acknowledgments}
We gratefully acknowledge discussions with Andreas Beyer,  Daniel Barker, Mathieu Cl\'ement-Ziza, Sin\'ead Collins, and Michael Nothnagel. This work was supported by the DFG under SFB 680. \\

{\centering
\section*{\large Appendix I: Short-time dynamics}
}

The statistics of states~(\ref{eq:nlinesselectiveall}) was derived in the limit of long evolutionary times (equilibrium). In general, the statistics of states depends on the length of branches of the phylogenetic tree (which we assume to be known). In this appendix, we derive the statistics of states in the limit of short evolutionary times and derive the corresponding log-likelihood score. At short evolutionary times, at most one mutation changing the state has fixed at each locus and across the phylogeny. 

Again we consider loci that are monomorphic in each population and identical initially. Then a mutation appears in one population and (with a certain probability) is fixed. The fixation probability depends on fitness, so the relative frequencies of such events at different loci allow in principle the inference of selection. Such short evolutionary times are characterized by $n \mu t \ll 1$, nevertheless the total number of diverged loci, characterized by $n \mu t L$ (where $n$ is the number of lines and $L$ is the total number of mutable loci affecting the trait), must still be at least of order one. Since our observable is the relative number of times mutations have fixed in one particular line (relative to other lines), the total number of mutable loci does not enter the statistics of states. In the regime of short evolutionary times, the ancestral states of the loci and the phylogeny of the lines affect the statistics of states, so general results for $n$ lines are unwieldy. Here, we compare the cases of $n=2$ and $n=3$. 

We start with the case of two lines, and consider a locus where one line has undergone a single change of state since the last common ancestor. This change can occur in either line, the relative probabilities for the change to occur in a particular line equal the relative rates at which the transition between states occurs in the two lines.  The transition rates between states (substitution rates) in a given line are $\mu_{+}\frac{4Ns_i a}{1-e^{-4Ns_i a}}$ and $\mu_{-}\frac{-4Ns_i a}{1-e^{4Ns_i a}}$ for the transition from $c=-$ to $q_{i}=+$ and $c=+$ to $q_i=-$, respectively (\citen{Kimura1962}) (the factor $4$ comes about because the phenotype changes by $2a$ during the transition). In general, the mutation rates $\mu_{+}$ (from $-$ to $+$) and $\mu_{-}$ ($+$ to $-$) will be different. Yet, for the relative probabilities of a mutation in one of the lines given the ancestral state a difference in mutation rates does not play a role as both lines start with the same ancestral state. This leads to the 
probability for the transition to occur in line $i$ 
\begin{equation}
P(i|a,c)=\frac{s_c(a,Ns_i)}{s_c(a,Ns_1) + s_c(a,Ns_2)} \ ,
\end{equation}
where we define the shorthand $s_c(a,Ns)=  \frac{-4Nsca}{1-e^{4Nsca}}$. $s_{i}$ ($i=1,2$) is the selection strength on the trait in line $i$.
Given two lines, both final configurations $(q_1,q_2)=(+-)$ and $(-+)$ can be reached from either ancestor $c=\pm$.\\

If the ancestral states are unknown, one can average over both possible ancestors. Writing the probability of ancestral state $c$ as $P(c)=e^{c Ns_{\text{anc}} a}$ and relative rates as $s_c(a,Ns)$, the dependence on the multiplicity parameter drops out again and we obtain  
\begin{equation}
P(q_1,q_2|a)=\frac{P(q_2)  s_{q_2}(a,Ns_1) + P(q_1) s_{q_1}(a,Ns_2)}{\sum\limits_{c=\pm 1} \sum\limits_{i=1}^2 P(c)  s_c(g,Ns_i)} \ .
\label{eq:twolinesshortnoanc}
\end{equation}
We have assumed that the distribution of states in the ancestral line has reached equilibrium under some selection strength $s_{\text{anc}}$, which will be inferred by maximum likelihood. 

\begin{figure}[htbp]
\begin{center}
\includegraphics[width=0.35\textwidth]{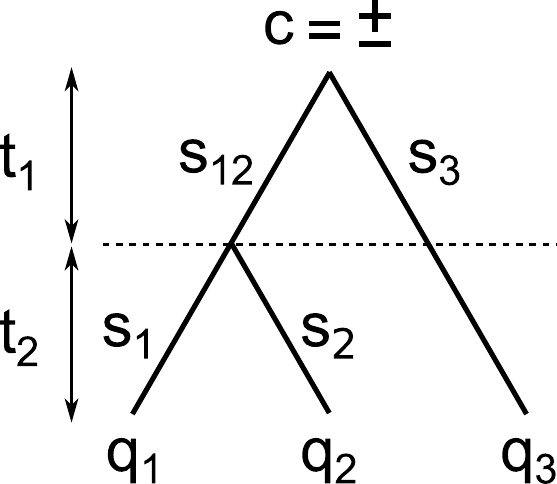}
\caption{\small {\bf Phylogenetic tree for three lines.} In the short-time limit, the states $(q_1,q_2,q_3)$ which can be reached by a single mutation from an ancestral state $c$ depend on the phylogenetic tree. The branch lengths $t_1$ and $t_2$ and selection strengths determine the relative mutation probabilities in the different branches.}
\label{fig:PhylogeneticTree}
\end{center}
\end{figure}

Considering three lines, four of the six possible diverged configurations can be assigned a unique ancestor: Denoting line 3 as the outgroup (see Figure~\ref{fig:PhylogeneticTree}), configurations $(q_1,q_2,q_3)=(+--)$ and $(-+-)$ diverged from the ancestral state $c=-$, configurations $(-++)$ and $(+-+)$ from ancestor $c=+$. Configurations $(++-)$ and $(--+)$ can either be reached by a mutation in the ancestor of lines 1 and 2 or a mutation in line 3. 
One can write the relative probabilities of the $6$ state configurations excluding $q_1=q_2=q_3$ as 
\begin{align}
\label{eq:short_probs}
--+ \qquad & (t_1 + t_2) P(-) s'_-(a,Ns_3) + t_1 P(+) s_+(a,Ns_{12}) \nonumber\\[0.1cm]
-+- \qquad & t_2 P(-) s_-(a,Ns_2) \nonumber\\[0.1cm]
+-- \qquad & t_2 P(-) s_-(a,Ns_1) \nonumber\\[0.1cm]
++- \qquad & t_1 P(-) s_-(a,Ns_{12}) + (t_1 + t_2) P(+) s_+(a,Ns_3) \nonumber\\[0.1cm]
+-+ \qquad & t_2 P(+) s_+(a,Ns_2) \nonumber\\[0.1cm]
-++ \qquad & t_2 P(+) s_+(a,Ns_1)\,,
\end{align}
where the times $t_1$ and $t_2$ account for the different branch lengths of the phylogenetic tree (see Figure~\ref{fig:PhylogeneticTree}). With these relative probabilities, the statistics of states in the three lines is
\begin{align}
Q_s(q_1,q_2,q_3|a)= &\frac{1}{Z} \Big(t_2 P(k) s_k(a,Ns_{q=-k}) + \delta_{q_1,q_2} t_1 \cdot \nonumber \\
&[P(k) s_k(a,Ns_3) + P(-k) s_{-k}(a,Ns_{12})]\Big),
\label{shorttimestatistics}
\end{align}
where we define the shorthand $k=q_1+q_2+q_3=\pm 1$, and $Ns_{q=-k}$ denotes the selection strength of the line with the minority state (e.g. $Ns_3$ for the configuration $(--+)$) and $Z=\sum_{q_1,q_2,q_3=\pm 1}' Q_s(q_1,q_2,q_3|a) $. Again, the two states with $q_1=q_2=q_3$ are excluded from this sum.

\begin{figure}[htbp]
\begin{center}
\includegraphics[width=0.55\textwidth]{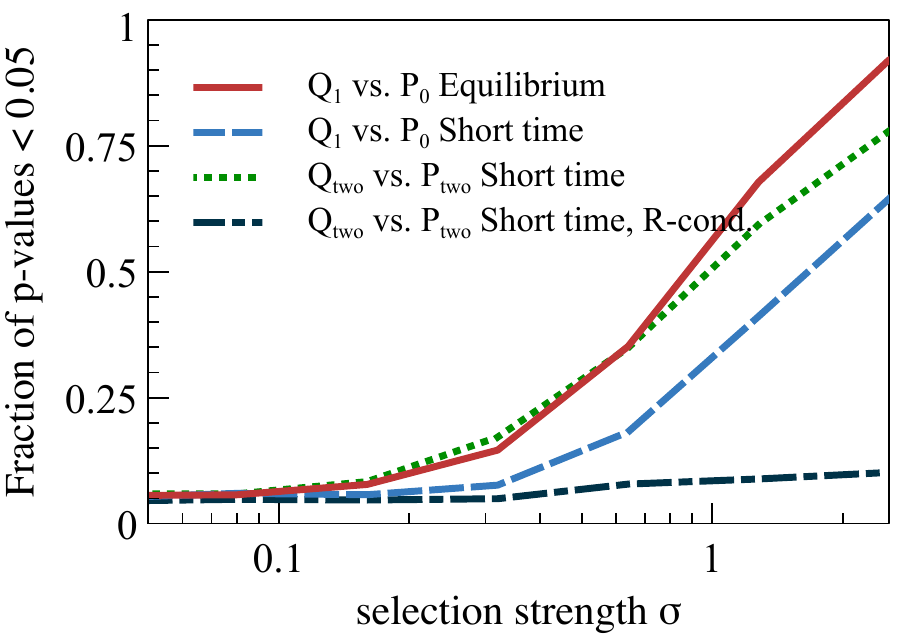}
\caption{\small {\bf Statistical significance of the tests for short evolutionary times.} In three lines, the selection test for short evolutionary times applied to artificial data created in the short-time limit shows less statistical power than the equilibrium test applied to data generated at long evolutionary times, but still allows to identify selection in a reasonable parameter range. However, for two lines under conditioning on $R_{\text{max}}$ the short-time test barely has any statistical power, analogously to the equilibrium case, where it has none. Parameters: The phylogenetic branch lengths $t_1$ and $t_2$ are taken equal to each other. The other parameters are as in Figure \ref{fig:OddsScore}.}
\label{fig:Simulations_ShortTime}
\end{center}
\end{figure}

Analogous to the equilibrium case, the statistics of states~(\ref{shorttimestatistics}) for different hypotheses $P_0$ and $Q_1$ etc. enters a log-likelihood score of the form (\ref{eq:LogScore}). To compare the resulting tests under different evolutionary scenarios, we perform numerical simulations at short evolutionary times as in section~\textit{Testing for selection at different evolutionary times}. No knowledge of the ancestral states is assumed.
Under the selective scenario $Q_1$, we find that the statistical power of the short-time test on three lines on short-time data is somewhat lower than the three-line equilibrium test applied to data for long evolutionary times at the same number of diverged loci (see Figure \ref{fig:Simulations_ShortTime}), but still allows to detect selection. On the other hand, for two lines the test under conditioning on $R_{\text{max}}$ gives hardly any significant results (see Figure \ref{fig:Simulations_ShortTime}), while the $R_{\text{max}}$-conditioning for three lines as well as the Holm--Bonferroni correction for two and three lines result allow to infer selection in a reasonable parameter range.\\

{\centering
\section*{\large Appendix II: Pedagogical example for the maximum entropy principle}
}

\begin{figure}[htbp]
\begin{center}
\includegraphics[width=0.65\textwidth]{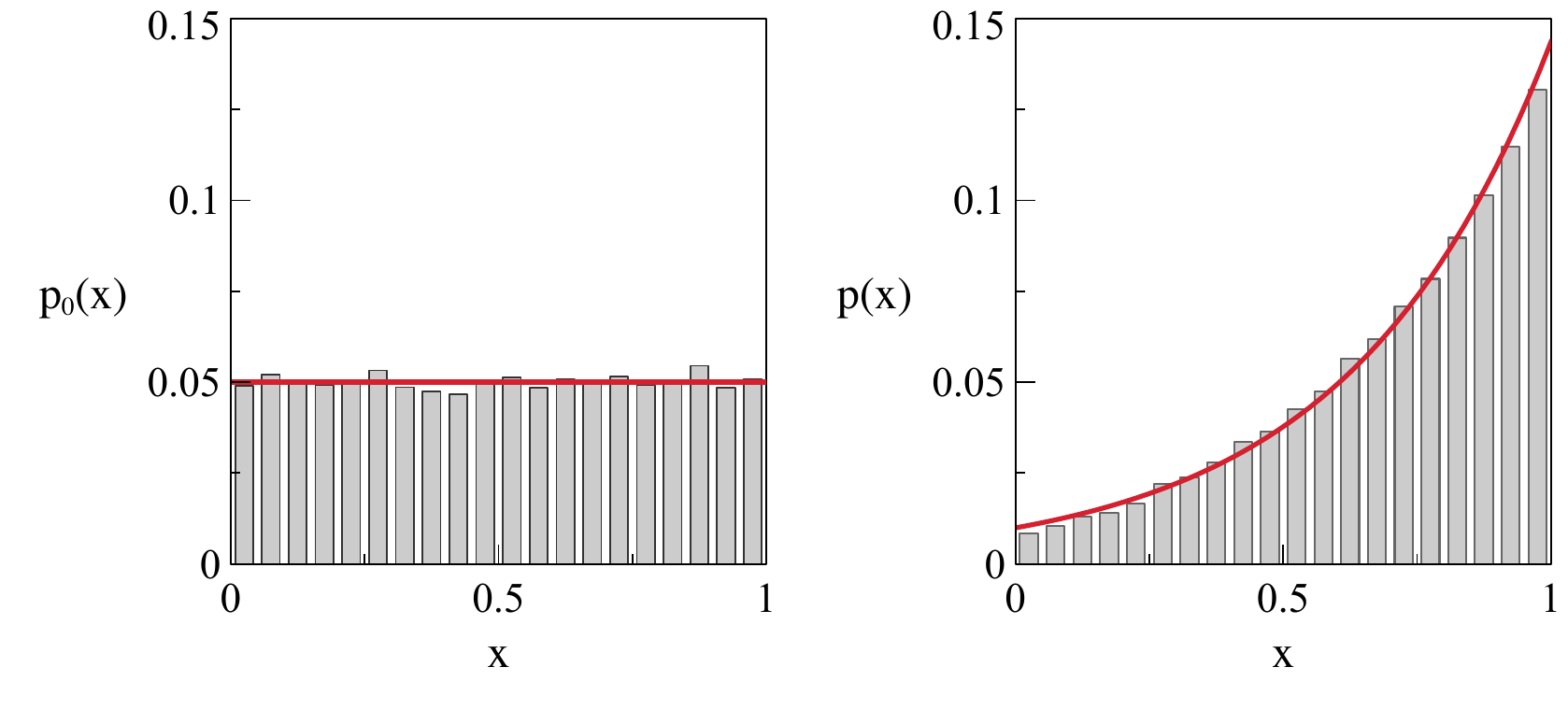}
\caption{\small {\bf A biased distribution can be inferred with the maximum entropy method.} Ascertainment leads to a biased distribution, which is derived using the maximum entropy method. Left: Histogram for 1000 sets of ten random numbers each drawn from a uniform distribution (red line) in the interval [0,1]. Right: Only sets of numbers are retained which have a sum $S$ close to $m=8$ ($7.95<S<8.05$). In these sets higher numbers appear more often than in the uniform distribution. The biased distribution takes on an exponential form given by the maximum-entropy distribution (\ref{eq:maxentrsolution}) (red line).}
\label{fig.PaedagogicalExample}
\end{center}
\end{figure}

Here, we give a simple concrete example to illustrate the link between ascertainment bias and the maximum entropy principle. Consider a uniform distribution $p(x)$ on the interval $[0,1]$, from which ten numbers are drawn independently (see Figure~\ref{fig.PaedagogicalExample} left). If one repeatedly draws such sets of ten numbers, the sum over each set will fluctuate from set to set with a mean value of $5$. In the next step, we only retain those sets whose sum is close to some value of $m\neq 5$. The numbers in these sets follow a non-uniform distribution and for $m>5$ one finds that larger values $x$ appear with a higher probability compared to the uniform distribution (see Figure~\ref{fig.PaedagogicalExample} right). Although each of these numbers was originally drawn from the uniform distribution, retention of sets with a particular mean value introduces a bias in the observed distribution of $x$. This is the ascertainment bias induced by conditioning the sum of each set. The principle of maximum entropy allows to determine the exact form of this biased distribution $p(x)$. We maximize the relative information entropy between the distribution $p(x)$ and the original (uniform) distribution $p_0(x) = 1$ for $x\in[0,1]$
\begin{equation}
H(p) = -\int\limits_0^1 \dx\, p(x) \log \frac{p(x)}{p_0(x)} \,,
\label{eq.information_entropy}
\end{equation}
subject to the constraints 
\begin{equation}
\int\limits_0^1 \dx\, p(x)  = 1\,, \qquad \int\limits_0^1 \dx\, x p(x)  = \frac{m}{N}\,,
\label{eq.constraints}
\end{equation}
where $N=10$ is the size of each set. Here, the first constraint ensures the normalization of $p(x)$ and the second constraint fixes the mean value of $x$ to $m/N$. 
Introducing Lagrange multipliers to maximize (\ref{eq.information_entropy}) subject to the constraints (\ref{eq.constraints}) leads to (\citen{ETJaynes1957}) 
\begin{align}
&-\int\limits_0^1 \dx\, p(x) \log p(x)  + \lambda_1 \left(\int\limits_0^1 \dx\, p(x) - 1\right) \nonumber\\
&+ \lambda_2 \left(\int\limits_0^1 \dx\, x p(x) - m/N\right)
\label{eq:maxentreq}
\end{align}
to be maximized with respect to $p(x)$. 
Differentiating (\ref{eq:maxentreq}) with respect to $p$ and setting the derivative to zero gives
\begin{equation}
p(x) = e^{\lambda_2 x + \lambda_1-1}.
\label{eq:maxentrsolution}
\end{equation}
Ascertainment bias thus makes $x$ exponentially rather than uniformly distributed, with coefficients $\lambda_1$ and $\lambda_2$ determined by the constraints (\ref{eq.constraints}). For $m=8$ and $N=10$ one obtains $\lambda_1 \approx -1.62$ and $\lambda_2 \approx 2.67$; the result for $p(x)$ shown in Figure~\ref{fig.PaedagogicalExample} agrees perfectly with the histogram of numbers in sets with a constrained sum.

Suppose one did not know whether the original distribution $p(x)$ from which the data were drawn was uniform or not and one had access only to data subject to the known constraint. If the distribution of the empirical data deviates from or agrees with the maximum entropy distribution $p(x)$, then this deviation or agreement could be used to quantify the likelihood that the original data came from the uniform distribution (vs. an alternative hypothesis). We follow the analogous approach with the score~(\ref{eq:LogScoreh}) to tell whether a particular statistics of states more likely comes from neutral evolution in combination with ascertainment bias (vs. an alternative scenario involving selection). \\

Finally, we sketch the derivation of the equilibrium statistics of states $P(q)$, which also follows an exponential form~(\citen{Iwasa1988};\,\citen{Berg2004};\,\citen{Sella2005}). For a finite population evolving under genetic drift and selection at low mutations rates, Kimura~(\citen{Kimura1962}) gives the rate at which a mutation appears and spreads to fixation as $u_{\Delta F}=\mu N \frac{1-\exp\{-2 \Delta F\}}{1-\exp\{-2 N\Delta F\}}$, where $\Delta F$ is the fitness difference relative to the pre-existing allele and $\mu$ the mutation rate. This rate obeys an exact relationship for forward and backward mutations $u_{\Delta F}/u_{-\Delta F}=\exp\{2 (N-1) \Delta F\}$ (detailed balance). Approximating $N-1$ by $N$, the equilibrium distribution over alleles is then $\sim \exp\{2 N F\}$ ~(\citen{vanKampen}), where $F$ is the fitness function of alleles.  Grouping together alleles corresponding to the same state of a locus yields~(\ref{eq:KimuraOhta}). When $F$ is linear in the states of loci, the corresponding probability distribution factorizes of loci.

\bibliographystyle{genetics.bst}
\renewcommand\refname{Literature Cited} 
\bibliography{Citations}

\end{document}